\documentclass{article}
\usepackage{graphicx} 
\usepackage{amsmath}
\usepackage{amssymb} 
\usepackage{braket}
\usepackage[linesnumbered,ruled,vlined]{algorithm2e}
\usepackage{threeparttable} 
\usepackage{url}
\usepackage{float}
\usepackage{empheq} 
\title{Variational free complement method with Gaussian expanded complement functions: extensions of general atomic framework including helium and lithium ground states with geminal induced expansion length}
\author{Cong Wang \thanks{PO Box 26 Okemos, MI, 48805 (USA); congwang.webmail@gmail.com}}
\date{}

\newcommand{\G}{\mathrm{G}}
\newcommand{\s}{\mathrm{s}}
\newcommand{\orb}{\mathrm{orb}}
\newcommand{\gem}{\mathrm{gem}}
\newcommand{\power}{\mathrm{power}}
\newcommand{\pair}{\mathrm{pair}}
\newcommand{\en}{\mathrm{en}}
\newcommand{\ee}{\mathrm{ee}}
\newcommand{\unique}{\mathrm{unique}}
\newcommand{\before}{\mathrm{before}}
\newcommand{\after}{\mathrm{after}}

\begin{document}

\maketitle

\begin{abstract}
We expand the previous framework of hierarchical-decontraction Gaussian expanded complement functions approach (arXiv: 2603.16262v2) of the free complement (FC) method to $N$-electron atomic systems including general permutations.
We consider the possibility that the Gaussian expansion length depends on the presence of the geminal $g$ function. To be more specific, fewer expansion length shall be used when the two-electron $g$ function is combined with the initial wavefunction. The present approach is applied to the ground states of the helium and lithium atoms. The complement function generations via 2s$\rightarrow$1s channel in the lithium ground state are discussed. Symmetries for the associated $N$-electron integrals are analyzed.

\end{abstract}

\section{Introduction} \label{sec_intro}
Numerically exact solutions of quantum many-electron systems can provide a wide range of accurate predictions, reference values of computational approaches \cite{shiozaki2009higher,Valeev_PRF_2009_46811G6,bubin2013born,mitroy2013theory,jiang2025neural}, and data for machine learning \cite{slootman2024accurate,scherbela2024towards}. The FC theory \cite{nakatsuji2004scaled,nakatsuji2005general,nakatsuji2012discovery} is a framework to these goals. 

We have introduced the Gaussian expanded complement approaches to circumvent the integral difficulties \cite{wang2025variational} and the hierarchical decontraction to mitigate the exponential complexities of the number of the complement functions \cite{wang2026free}. The energy convergence has been analyzed \cite{wang2026variational}. These approaches are free from stochastic fluctuations of the sampling methods \cite{jiang2025neural} and can provide upper and lower bounds of the exact energies \cite{nakashima2008accurately,Ireland2021IntegralsLowerBounds,ronto2023lower,nakatsuji2024accurate}. 

Extending the hierarchical decontraction approach  would introduce multiple loops in computing the matrix elements, when there is no one-electron $g$ function multiplied on the initial wavefunction nor simplifications from the L\"{o}wdin-type rules \cite{lowdin1955quantum_a,lowdin1955quantum_b,nakashima2020free,wang2026free}. This complexity may be circumvented by the $s_{ij}$ approach \cite{nakatsuji2020solving}. The rate of convergence may be expected to be similar to the conventional configuration interaction method \cite{sims2006high}, especially in the scale more accurate than 1 milli-Hartree total energy accuracy \cite{prendergast2001impact}.

As noticed from the energy-based selection in the previous work \cite{wang2026free}, with the presence of the geminal $g$ function, a smaller Gaussian expanded length would be needed to reach the similar accuracy. This advantage may be realized by an efficient energy selection algorithm. To reduce the computational cost at the early state prior to the overlap selection \cite{von2008trapped,von2009correlated,rakshit2012hyperspherical,mitroy2013theory,Kalaee2014,Mosegaard2018,moriya2023novel,coomar2022quantum,wang2025variational,wang2026free}, we shall choose the Gaussian expansion length according to the presence of the geminal $g$ function.

At the practical side, although helium ground state encodes many insights of electron correlation, advancing theoretical approaches on helium atom prior to expand the software infrastructure for $N$-electron ($N \geq 3)$ systems may bring increasing amount of efforts in applying the approaches to more-electron systems. This is largely due to the symmetric spatial wavefunction of the helium ground state relatively special. It does not appear in more-electron systems. Thus, it is preferable to enhance the implementations into general system, prior to implement additional methodologies. 

To this end, the objective of the present work are two folds: \begin{itemize}
    \item [(i)] expand the framework into $N$-electron system ($N>2$);
    \item [(ii)] introduce the Gaussian expansion length depending on the presence of geminal function;
\end{itemize}
Furthermore, we notice for more-electron system, such as lithium ground state, additional complement function from the potential operator as 2s$\rightarrow$1s will appear. This seems have not been discussed in the earlier work \cite{nakatsuji2024accurate}.  Some additional issue for matching indices in deciding contraction/decontraction also appears via this channel.  It is better to describe them in the theory and algorithms.

Based on the aforementioned considerations, the  present work is organized as following. In Section \ref{formulae}, we describe the hierarchical decontraction framework for general $N$-electron system and the various expansion lengths with and without the geminal function, called the geminal induced expansion length.  The algorithms for eliminating duplicated complement functions, generating complement functions with $g$ functions, expanding into product of Slater functions, forming Gaussian expanded complement functions, overlap matrix selection, energy computations, and symmetries of related integrals are described in the subsections.  
 
In Section \ref{implementation}, the main software and libraries used in the implementation of the present algorithms are described. 
In Section \ref{result}, the numerical results for the helium and the lithium ground states are presented. In Section \ref{summary_outlook}, summary and outlook are provided.  
In Appendix, a comparison between different orderings of the complement functions is provided. Alternative decontraction and contraction schemes are discussed.  In Supplementary Material, we provide an example including lists of complement functions and the ground state energy of the lithium ground state. This may facilitate a reader to reproduce the results of the present work.

\section{Methodology} \label{formulae}
\subsection{Geminal induced expansion length and the theoretical context} \label{outline}
The FC wavefunction has the functional form in the simplest iterative configuration interaction (SICI) approach \cite{nakatsuji2004scaled,nakatsuji2005general,nakatsuji2012discovery,nakatsuji2018solving,nakatsuji2019solving}
\begin{align}
   \psi_n = \prod_{m=0}^{n-1} \left[ 1 + C_m g (H - E_m) \right] \psi_0  
\end{align}
where $n$, $C_m$, $g$, $H$, $E_m$, and $\psi_0$ are the expansion order, the expansion coefficient, scaling function,  Hamiltonian of the system, an energy for the generation process, and the initial wavefunction, respectively. It is possible to only use the potential operator in the Hamiltonian in the FC method, known as the p-alone approach \cite{nakatsuji2022accurate}.

The coefficient $C_m$ can be generalized to a number of $\mathcal{I}$ components, as $C_m^\mathcal I$ associated to the partition of the Hamiltonian $H = \sum_{\mathcal{I=1}}^{N_D} H_{\mathcal{I}}$, in the ICI with $N_D$ variables (ICIND) approach \cite{nakatsuji2001structure,nakatsuji2004scaled,nakatsuji2005general}. In the FC approach, all linear independent functions, as the complement functions, are associated to the linear independent expansion coefficients $\{c_i \}$ \cite{nakashima2018solving}
\begin{align}
    \psi_n = \sum_{I=1}^{M} c_I \phi_I
\end{align}
where $M^\before$ \cite{nakatsuji2012discovery} is the number of the linear independent functions $\{ \phi_i \}$ \cite{nakatsuji2004scaled,nakatsuji2005general,nakashima2007solving, nakatsuji2012discovery}.  Notice some linear combinations can used in the basis functions. For example, 
\begin{itemize}
    \item[(i)] $1 + P_{12} $ in $(1 + P_{12} ) g_1^{n_1} g_2^{n_2} g_{12}^{n_{12}} e^{-\zeta(r_1 + r_2)}$ of the helium ground state spatial wavefunction \cite{nakatsuji2024accurate};
    \item[(ii)] for the simultaneous spin eigenfunction  $\{S^2, S_z \}$: $\chi_+ \chi_- - \chi_- \chi_+$ \cite{nakatsuji2022potential}, where  $\chi_\pm$ is the  eigenfunction of the one-electron $s_z$ operator with eigenvalue $\pm 1/2$, respectively. The atomic units (a.u.) are used in this work by default;
    \item[(iii)] for the initial wavefunction, $(1-r+r^2) e^{-r/3}$ \cite{kurokawa2019solving,kurokawa2020solving} and contracted terms in the hierarchical scheme \cite{wang2026free};
    \item[(iv)] for the $g$ function and the generated complement functions: $1-e^{-\gamma r}$ type of  $g$ function \cite{nakatsuji2022accurate,nakatsuji2024accurate,nakatsuji2024exact} can generate $(1-e^{-\gamma r})/r$ expression via the potential operator ($g/r$ type terms \cite{nakatsuji2024accurate}). For the 1s-type initial wavefunction,  this type of terms contain individually diverged expression \cite{nakatsuji2004scaled}. The combined form leads to finite matrix elements. These terms have severe linear dependence and inefficient energy improvement. They are not used in standard FC calculations \cite{nakatsuji2024accurate} nor the previous works \cite{wang2025variational, wang2026free, wang2026variational}.
\end{itemize}

We consider an atomic electronic initial wavefunction, 
\begin{align}
      \psi_0 &= \mathcal A (\Phi_0 \Theta) \\
      \Phi_0 &= \prod_{i}  r_{i}^{\mu_{i}}  Y_{l_{i} m_{i} }(\theta_{i}, \varphi_{i} )  e^{-\zeta_{i} r_{i} }   \label{psi0_general}
\end{align}
where $\mathcal A := \sum_p (-1)^p P$ is the unnormalized antisymmetrizer \cite[p. 78]{pauncz2012construction}\cite{fischer2014vertically}. $\Phi_0$ and $\Theta$ are the spatial part and $\{S^2, S_z \}$ simultaneous eigenfunction \cite[p. 284]{maruani2003new}, respectively. Also, $\{S^2, S_z \}$ form simultaneous eigenstates with the many-electron non-relativistic Hamiltonian and antisymmetries of electrons \cite[p. 613]{liboff2003introductory} \cite[p. 44]{Fayer2018}. $\mu_{i}$ is the power of electron-nucleus radial function in the initial wavefunction.   The spherical harmonics function is in the form of uncoupled basis at the present stage. It is of interest to consider coupled basis  \cite{sims2009hylleraas,mitroy2013theory} expression to better incorporate symmetry properties of the many-electron Hamiltonian \cite[p. 1016]{cohenvol2} \cite{mitroy2013theory}. For S-state via $Y_{00}$ orbitals, this is automatically satisfied.  $r_{12}^{n_{12}}$ can be viewed as the scalar component of higher angular moment products \cite[p. 161 - 164]{varshalovich1988quantum} \cite{sack1964generalization} \cite[Remark V. 3.1.5]{Morel2019}.

In Eq. \eqref{psi0_general}, one eigenfunction of $\{S^2, S_z\}$, $\Theta$, is used.  There are two such eigenfunctions for the lithium ground state \cite{Larsson1968,CencekRychlewski1993,YanDrake1995}.  As derived in Refs. \cite{Larsson1968,CencekRychlewski1993} and numerically demonstrated in Ref. \cite{YanDrake1995}, one single $\{S^2, S_z \}$ eigenfunction in Eq. \eqref{psi_form_li} is sufficient to obtain the ground state energy of the lithium atom. 

Notice for  $\mu_{i} > 0$, the potential operator can lower $\mu_{i}$ to a non-negative number. The resulting term can be formally written as $g/r$ as in the previous study \cite{nakatsuji2024accurate}.

Nevertheless, for the lithium ground state, the 2s Slater-type initial orbital with the potential operator multiplied on it once, no individual diverged term is inside as long as the associated power of $r$ is positive semi-definite. This is to the contrary of the helium ground state  \cite{nakatsuji2024accurate}.  
Since 2s and 1s orbitals typically have different exponents, no violation of the Pauli exclusion principle. Hence, with $g = \sum_{i} g_{i} + \sum_{i<j} g_{ij}$ \cite{nakashima2018solving} and $1-e^{-\gamma r}$ type $g$ function, we have the form of the complement function at order $n$:
\begin{align}
 \phi_I &= \mathcal A (\Phi_I \Theta)    \\   
\Phi_I &=  \prod_{j<k}  g^{n_{jk}}_{jk}   \prod_{i}    g^{n_{i}}_{i}  V^{\nu_{i}}_{i}    \psi_0,   \,\,\,    0 \leq \sum_{i} n_{i} + \sum_{j<k}  n_{jk} \leq n, 0 \leq \sum_{i} \nu_{i} \leq n, \nu_{i} \leq \mu_{i}   \label{g_psi0}
\end{align}

Combining \eqref{g_psi0} and \eqref{psi0_general} leads to a series of products of one- and two-electron Slater functions, schematically as $(1-e^{-\gamma r})^n \cdots$. This allows us to reorganize the complement function at order $\{ n_{i}, n_{jk}  \}$ by expanding $(1-e^{-\gamma r})^n$ as  
\begin{align}
  \mathcal A \left\{  \left[ \prod_{j<k}     e^{- n_{jk} \gamma_{jk}}  \prod_{i}  r_{i}^{\mu_{i} - \nu_{i}}  Y_{l_{i} m_{i} }(\theta_{i}, \varphi_{i} )   e^{- (n_{i} \gamma_{i} + \zeta_{i}) r_{i}}   \right]   \Theta \right\} \label{g_psi0_expand}
\end{align}

In the hierarchical decontraction scheme \cite{wang2026free}, the contraction/decontraction is made without/with the $e^{-\gamma r}$ part of the $g$ function. The presence of $\mu_{i} - \nu_i$ forms an addition-type distinction in the power part of the Slater functions than the helium ground state in the previous work \cite{wang2026free}. Hence, we propose \cite{wang2026free}
\begin{align}
 r_{}^{\tilde \mu} e^{- \tilde \zeta r_{} } Y_{lm} &\rightarrow \{ e^{-\alpha_{\tilde \mu lm}^{( p)} r_{}^2}  Y_{lm} | p=1,\cdots, n_{\G} \},   \quad
\begin{aligned}[t]
  &  \tilde \mu = \mu - \nu,  \nu = 1,2,\cdots, \mu, \\
  &  \textrm{ or }  \tilde \zeta = \zeta  + k \gamma, k=1,2,\cdots \label{choice_decontraction}
\end{aligned} \\
 r_{}^{\tilde \mu} e^{-\tilde  \zeta r_{} }  Y_{lm}  &\rightarrow   \sum_{p=1}^{n_{\G}} c_{ \tilde \mu lm}^{(p)} e^{-\alpha_{\tilde \mu lm}^{(p)} r_{}^2 }    Y_{l_{} m_{}},  \quad \tilde \mu = \mu  \textrm{ and } \tilde \zeta = \zeta   
\label{choice_contraction}
\end{align}

Here $\tilde \zeta$ and $\zeta$ are generally denoted to the exponents associated to electron-nucleus and electron-electron distances, similar to $r$.  $\mu$ and $\zeta$ correspond the exponent and power in the initial wavefunction Eq. \eqref{psi0_general}.

Notice there can be some ambiguity in contraction/decontraction, since the conditions in \eqref{choice_decontraction} and \eqref{choice_contraction} only specifies spatial information. The spatial functions are a series of permutations by the antisymmetrizer. For two linear-independent lithium wavefunction, $\mathcal A(\Phi_0 \Theta)$ and $\mathcal A(\Phi_1 \Theta)$,  with $\Phi_0 = e^{-r_1 -2r_2 - 3r_3}, \Phi_1 = e^{-3r_1 -2r_2 - r_3}, \Theta =  \left[ \chi_+ (1) \chi_-(2)  - \chi_- (1) \chi_+(2) \right] \chi_+ (3) $, $\mathcal A = 1 - P\left[\begin{smallmatrix} 1& 2 \\ 2 & 1\end{smallmatrix}\right] - P\left[\begin{smallmatrix} 1& 3 \\ 3 & 1\end{smallmatrix}\right]  - P\left[\begin{smallmatrix} 2& 3 \\ 2 & 2\end{smallmatrix}\right]  + P\left[\begin{smallmatrix} 1& 2 & 3 \\ 2 & 3 & 1\end{smallmatrix}\right] + P\left[\begin{smallmatrix} 1& 2 & 3 \\ 3 &  1 & 2 \end{smallmatrix}\right]$ \cite{frolov2011atomic}, the two rows symbols, e.g.  $\left[\begin{smallmatrix} 1& 2 & 3 \\ 2 & 3 & 1\end{smallmatrix}\right]$, means permutation \cite[26.13]{NIST:DLMF_1_2_8}. The linear independency is discussed in Subsection \ref{unique}. Here $\Phi_1$ can be obtained by some flexible form of $g$ function to modify its exponents.  Permutation $P_{13}$ will let $\Phi_1 $ have the same spatial structure as $\Phi_0$.

However, the choice of decontraction/contraction is to form a hierarchy of corrections of the initial wavefunction, no additional permutation than the initial wavefunction would be expected for the comparison. Thus, $ \Phi_1 = e^{-3r_1 -2r_2 - r_3}$ would be the form for choosing the decontractions.

In addition, the decontraction may only be made via $\tilde \zeta$, not $\tilde \mu$ in \eqref{choice_decontraction} and \eqref{choice_contraction}. We shall discuss these options in Appendix of the present work.

Though the aforementioned description is for atomic case. The generalizations to molecular system is expected to be straightforward, since the FC methods have been applied to molecular systems \cite{kurokawa2020solving}.

From expression \eqref{g_psi0_expand}, we can label the Gaussian expansion length as $n_{\G q}  $. Here $q=1,2,3,\cdots, 12,13,\cdots$ corresponds to the index of the expanded variable. In principle,  $n_{\G q}  $ can be a function of any computational setup, including the descending/ascending orders for the complement functions \cite{wang2025variational,wang2026free}. In the present scope, we consider the expansion length as a function of the powers of the Slater functions, $n_1, n_2, \cdots, n_{12}, \cdots$. We denoted it as $ n_{\G q} (\{ n_i, n_{ij} \} ) $.

Hence, the geminal induced expansion length approach in the present scope is 
\begin{align}
  \left. n_{\G q} (\{ n_i, n_{ij} \}) \right|_{n_{ij} = 0 } \geq    n_{\G q} (\{ n_i, n_{ij} \}) |_{n_{ij} > 0 } \label{inequal_ng}
\end{align}

In addition, it is possible to consider orbital induced expansion length. Namely, we replace $n_{ij} = 0$ and $n_{ij} > 0$ to $n_{i} = 0$ and $n_{i} > 0$ in \eqref{inequal_ng}, respectively. There is some evidence support the possible effective of this approach \cite[Table 4, 1,0,0 row]{wang2026free}. Nevertheless, the overlap matrix based selection can remove some complement functions via the decontraction for the associated $i$-th electron. This likely makes the orbital induced expansion length less computational critical than the geminal induced expansion length. In the present work, we shall  only consider the geminal induced expansion length approach.

\subsection{Elimination of linear dependent complement functions} \label{unique}
Complement function \eqref{g_psi0_expand} could contain equivalent terms (difference up to a global prefactor). This can lead to linear dependency. For example, consider complement function in the lithium ground state
\begin{align}
& \mathcal A  \left[ r_3
e^{- (\zeta_{1\s} + n_{1\s} \gamma_{1\s} ) r_1 -  (\zeta_{1\s} + m_{1\s} \gamma_{1\s}  ) r_2 - \zeta_{2\s} r_{3} } \Theta  \right] \label{psi_form_li} \\
  \Theta   &=  \left[ \chi_+ (1) \chi_-(2)  - \chi_- (1) \chi_+(2) \right] \chi_+ (3)
\end{align}
As the result of antisymmetrization, $n_{1\s}=1, m_{1\s}=2$ is equivalent to $n_{1\s}=2, m_{1\s}=1$. 

To the contrary of the helium ground state, the spatial functions are not totally symmetric with respect to any permutation among electrons 1, 2, and 3 after the antisymmetrizations. Sorting the exponents \cite{wang2025variational} is no longer applicable to remove duplicate complement functions. Basis-function divisions and dropping-term conditions have been proposed in a similar context \cite{YanDrake1995, yan1998energies,puchalski2006ground,puchalski2008relativistic,PhysRevA.80.052507}. 

It is possible to remove this linear dependence effectively by the subsequent overlap selection \cite{von2008trapped,von2009correlated,rakshit2012hyperspherical,mitroy2013theory,Kalaee2014,Mosegaard2018,moriya2023novel,coomar2022quantum,wang2025variational,wang2026free}. To remove the duplicated complement functions at an earlier stage, it is also possible to compare the spatial and spin components after the antisymmetrizations for each generated complement functions.

Here we propose an algorithm to select linear independent complement functions within a single configuration state function, $\Theta$. In this algorithm, only the spatial components are used in the comparison.  This algorithm is related to the permutational structure of the matrix elements in the secular equation in Refs. \cite[Eq. (6)]{puchalski2006ground} and \cite[Eqs. (23) and (24)]{puchalski2008relativistic}. 

We first consider the matrix elements of the secular equation \cite{matyus2019pre}
\begin{align}
   O_{IJ} &= \braket{ \phi_I | \hat O | \phi_J } \\
   &=  \braket{ \mathcal A ( \Phi_I \Theta ) | \hat O | \mathcal A (\Phi_J \Theta) }  \\
   &= N! \braket{  \Phi_I \Theta  | \hat O | \mathcal A (\Phi_J \Theta) } \label{one_fold_permu}
\end{align}
where $\hat O = \hat H$ and $\hat 1$ correspond to the Hamiltonian and overlap matrices, respectively.  We have 
used $\mathcal A \mathcal A = N! \mathcal A$ \cite{reading1980k,hattig2009electronic}.  Here $\Phi_I$ and $\Phi_J$ are generally denoted to complement functions in the present work, not necessarily in the form of Eq. \eqref{g_psi0}. 

Eq. \eqref{one_fold_permu} suggests the linear dependency can be identified by comparing the expressions
\begin{align}
     \braket{   \Theta   | \mathcal A (\Phi_J \Theta) }  \label{permu_target}
\end{align}
Since the non-relativistic Hamiltonian operator $\hat H$ is spin-free,  $\braket{   \Theta   | \mathcal A (\Phi_{J'} \Theta) } \propto  \braket{   \Theta   | \mathcal A (\Phi_{J''} \Theta) } $
will lead to duplicated entries in the Hamiltonian and overlap matrices in the secular equation.

The aforementioned operational elimination can cover three levels of complement functions in the present work: 
\begin{itemize}
\item [(a)] Eq. \eqref{g_psi0} including $g$ functions and the associated powers;
\item [(b)] Slater functions from expanding the $g$ functions in the form of \eqref{g_psi0_expand};
\item [(c)] Gaussian expanded complement functions via Cartesian products of \eqref{choice_decontraction} and \eqref{choice_contraction}.
\end{itemize}
Notice only (c) is used in the secular equation and overlap selection. We can extend the operational elimination of duplication for the types (a) and (b), since they are also used in explicitly correlated energy calculations and lead to numerically exact results \cite{wang2012variational,nakatsuji2024accurate,nakatsuji2024exact}. Also see Ref. \cite{wang2012variational} for a more formal discussion.

The spatial complement functions are then presented by
\begin{equation}
\left(
\begin{aligned}
& n_{i}, \cdots, n_{jk}, \cdots  \\
& \gamma_{i}, \cdots, \gamma_{jk}, \cdots    \\
& \mu_{i} - \nu_i, \cdots,  0, \cdots  \\
& \zeta_{i}, \cdots,  0, \cdots  
\end{aligned}  
\right) \label{spatial_entries_g_function}
\end{equation}

\begin{equation}
\left(
\begin{aligned}
& \mu_{i} - \nu_i, \cdots,  0, \cdots  \\
& \zeta_{i}, \cdots,  \zeta_{ij}, \cdots  
\end{aligned}  
\right) \label{spatial_entries_slater}
\end{equation}
 
\begin{equation}
\left(
\begin{aligned}
& \sum_{p=1}^{n_\G} c^{(p)}_i \alpha_{i}^{(p)}, \cdots,  \alpha_{ij}, \cdots  
\end{aligned}  
\right) \label{spatial_entries_gauss}
\end{equation}
for types (a), (b), and (c), respectively. $n_\G=1$ in \eqref{spatial_entries_gauss} presents decontracted. For $\alpha_{ij}$ in \eqref{spatial_entries_gauss}, since the initial wavefunction, Eq. \eqref{psi0_general}, does not have geminal function, it is decontracted.

More specifically, in each row the order associated to $r_1, r_2, \cdots, r_N, r_{12}, r_{13}, \cdots, r_{1N}$, $r_{23}, \cdots, r_{2N}, r_{34}, \cdots$. In the third and fourth lines of \eqref{spatial_entries_g_function} and the first line of \eqref{spatial_entries_slater}, 0s are filled for the slots of the geminals.  

The algorithm of using spatial components to remove duplicated functions is presented in Algorithm \ref{alg_gen_unique}.

\begin{threeparttable}
\begin{algorithm}[H] 
    \caption{Generating non-duplicated complement functions for $N$-electron system} \label{alg_gen_unique}
    \KwIn{  $( \ket{ \Phi_I })$ that each  $\ket{ \Phi_I }$  is presented in  \eqref{spatial_entries_g_function} - \eqref{spatial_entries_gauss}, $\Theta$ in terms of the eigenfunction of one-electron operator $\hat s_z$, namely $\chi_\pm$, and \texttt{n\_electrons} of the system   
    }
    \KwOut{  $ (  \ket{  \Phi_I^{\unique} } )   $ }
    $ (  \ket{  \Phi_I^{\unique} } )   $  = () \;
     $(1,\cdots, N, 12, \cdots N-1 N ) \Leftarrow$ \, \texttt{n\_electrons} \;
    \For{  $ \ket{ \Phi_I } $   in     $( \ket{ \Phi_I })$       }{
     $( \,  (-1)^P \braket{  \Theta| P | \Theta }  \, )  \Leftarrow $ \,   $\Theta$  \;
    \texttt{combined\_permuted\_terms} = () \;

    \For{    $(-1)^P \braket{  \Theta| P | \Theta }$    in   $( \, (-1)^P \braket{  \Theta| P | \Theta }  \, ) $    }{      
     $(-1)^P \braket{   \Theta   | P| \Theta } \, \ket{\Phi_I \!\left[\begin{smallmatrix} 1& \cdots &N\\ \sigma_1 & \cdots & \sigma_N \end{smallmatrix}\right]}$    $\Leftarrow$  $(-1)^P \braket{  \Theta| P | \Theta }$ , $ \ket{ \Phi_I } $,  $(1,\cdots, N, 12, \cdots N-1 N )$ \;

     \eIf{  $\ket{\Phi_I \!\left[\begin{smallmatrix} 1& \cdots &N\\ \sigma_1 & \cdots & \sigma_N \end{smallmatrix}\right]} $ in  \upshape \texttt{combined\_permuted\_terms}  }
     { add  $(-1)^P \braket{  \Theta| P | \Theta }$ into the related term in  \upshape \texttt{combined\_permuted\_terms} }
     {    \texttt{combined\_permuted\_terms}  $\leftarrow$    $(-1)^P \!  \braket{   \Theta   | P| \Theta }$, \! $\ket{\Phi_I \!\left[\begin{smallmatrix} 1& \cdots &N\\ \sigma_1 & \cdots & \sigma_N \end{smallmatrix}\right]}$  \;  } 
    }

    \texttt{canonical\_permuted\_terms}       $\Leftarrow$ \texttt{combined\_permuted\_terms} 

    \If{ \upshape \texttt{canonical\_permuted\_terms}  in  $ (  \ket{  \Phi_I^{\unique} } )  $  }{continue}
       $ (  \ket{  \Phi_I^{\unique} } )   $    $\leftarrow$ \texttt{canonical\_permuted\_terms}
    }    
    \Return   $ (  \ket{  \Phi_I^{\unique} } )   $
\end{algorithm}
\end{threeparttable}

For the Algorithms in the present work, the following notations apply. $\Leftarrow$ and $\leftarrow$ correspond to assign value to the item at left and add at the end of the item at left, respectively. When it is self-evident not to present parenthesis in calculation, $\{\}$ and $()$ correspond to unordered set and ordered tuple, respectively. These notations also apply to Algorithms \ref{alg_gen_slater}, \ref{alg_gen_gauss}, and \ref{alg_sel} in Subsections \ref{gen_slater_function}, \ref{gen_gauss_function}, and \ref{overlap_sel_secular}, respectively. 
$\prec$ means order (ascending/descending, except $\prec_\mathrm{Slater}$ in Algorithm \ref{alg_gen_slater} which keeps the option of the software \texttt{SymPy} \cite{10.7717/peerj-cs.103}  order) of the associated entries.

Notice the concepts 'set' and 'tuple' in the Algorithms of the present work  are at the level of pseudo code. They do not necessarily correspond to the 'set' and 'tuple' data type, including mutability and orderability of the \texttt{Python} programming language \cite{van1991interactively}.

In line 2 of Algorithm \ref{alg_gen_unique}, we use the number of electrons form variables indices, $1,2,\cdots, N, 12, \cdots N-1 N$, these will be the objects of permutations in line 7 of Algorithm \ref{alg_gen_unique}.

In line 7 of Algorithm \ref{alg_gen_unique},  $\ket{\Phi_I \!\left[\begin{smallmatrix} 1& \cdots &N\\ \sigma_1 & \cdots & \sigma_N \end{smallmatrix}\right]} $, the indices under permutation $1,\cdots, N$ covers the indices of geminal functions, e.g., $12$ in $r_{12}$. $\sigma_i, i = 1,\cdots N$ as the permuted indices.

In Algorithm \ref{alg_gen_unique}, \texttt{combined\_permuted\_terms} stores entries in the format that each permutation of \eqref{spatial_entries_g_function} - \eqref{spatial_entries_gauss} associated with a weight.

Via line 8 of Algorithm \ref{alg_gen_unique}, the weight is the sum of numerically equal \eqref{spatial_entries_g_function} - \eqref{spatial_entries_gauss}.  Currently exact comparison for integer powers and $1\times 10^{-14}$ relative error, $|a-b|/\mathrm{max}(|a|,|b|)< 1\times 10^{-14}$ for numerical data \cite{python_math}, since the reported STO-$n$G basis set has at most about 13 significant figures \cite{stewart1970small,fernandez1988accurate}. In addition, \texttt{combined\_permuted\_terms} is ordered by insertions.

In line 12 of Algorithm \ref{alg_gen_unique}, the permutations are sorted in  lexicographical order (smaller first). For example, in case (a), the lexicographical order follows four rows, from upper to lower. First compare the row of $ n_{i}, \cdots, n_{jk}, \cdots $ for the lexicographical order. Then compare the second row for the lexicographical order. 

After that, the permutational coefficient of the leading term is normalized into 1,  dividing the weight of each term by the weight of the leading term.  If the weight of the leading term is zero, set the next term with non-zero weight as the leading term. If the weights of all permutations are zero, e.g., $(1s)^3$ configuration of lithium ground state, it is excluded.

In line 13 of Algorithm \ref{alg_gen_unique}, the thresholds are the same as line 8, besides exact match is used for the weights.

For example, the example in Subsection \ref{outline}, $\Phi_0 = e^{-r_1 -2r_2 - 3r_3}, \Phi_1 = e^{-3r_1 -2r_2 - r_3}$  are converted in line 12 as 
\begin{equation}
\begin{split}
\left(
\begin{aligned}
& 0,0,0,0,0,0 \\
& 1,2,3,0,0,0
\end{aligned}
\right)
\rightarrow\ & 1 \left(
\begin{aligned}
& 0,0,0,0,0,0 \\
& 1,2,3,0,0,0
\end{aligned}
\right)
-\frac{1}{2} \left(
\begin{aligned}
& 0,0,0,0,0,0 \\
& 1,3,2,0,0,0
\end{aligned}
\right) + 1 \left(
\begin{aligned}
& 0,0,0,0,0,0 \\
& 2,1,3,0,0,0
\end{aligned}
\right) \\
& + 1 \left(
\begin{aligned}
& 0,0,0,0,0,0 \\
& 2,3,1,0,0,0
\end{aligned}
\right) - \frac{1}{2} \left(
\begin{aligned}
& 0,0,0,0,0,0 \\
& 3,1,2,0,0,0
\end{aligned}  
\right)  - \frac{1}{2} \left(
\begin{aligned}
& 0,0,0,0,0,0 \\
& 3,2,1,0,0,0
\end{aligned}  
\right)
\end{split}
\label{cf_1}
\end{equation}

\begin{equation}
\begin{split}
\left(
\begin{aligned}
& 0,0,0,0,0,0 \\
& 3,2,1,0,0,0
\end{aligned}
\right)
\rightarrow\ & 1 \left(
\begin{aligned}
& 0,0,0,0,0,0 \\
& 1,2,3,0,0,0
\end{aligned}
\right)
+ 1 \left(
\begin{aligned}
& 0,0,0,0,0,0 \\
& 1,3,2,0,0,0
\end{aligned}
\right) + 1 \left(
\begin{aligned}
& 0,0,0,0,0,0 \\
& 2,1,3,0,0,0
\end{aligned}
\right) \\
& -2 \left(
\begin{aligned}
& 0,0,0,0,0,0 \\
& 2,3,1,0,0,0
\end{aligned}
\right) +1 \left(
\begin{aligned}
& 0,0,0,0,0,0 \\
& 3,1,2,0,0,0
\end{aligned}  
\right)  - 2 \left(
\begin{aligned}
& 0,0,0,0,0,0 \\
& 3,2,1,0,0,0
\end{aligned}  
\right)
\end{split}
\label{cf_2}
\end{equation}
The left-hand side (LHS) of \eqref{cf_1} and \eqref{cf_2} are the input and correspond to the complement function expression \eqref{spatial_entries_g_function}. The right-hand side (RHS) of  \eqref{cf_1} and \eqref{cf_2} correspond to  \eqref{permu_target}.

The difference of RHS in  \eqref{cf_1} and \eqref{cf_2}  suggest the LHS are not equivalent after the antisymmetrization. The current algorithm compare the $n$-th term, ($n=1,2\cdots$) at the RHS of \eqref{cf_1} and \eqref{cf_2}, one by one, for both coefficient and entries in the round brackets. If any comparison in the $n$-th pair between \eqref{cf_1} and \eqref{cf_2} is inconsistent, the result is inconsistent.  The numerical consistency is identified by numerical threshold, 

Since they are not duplicated, the LHS of \eqref{cf_1} and \eqref{cf_2} are the output.

The basis function dropping-term conditions  \cite{YanDrake1995, yan1998energies,puchalski2006ground,puchalski2008relativistic,PhysRevA.80.052507} can be regarded as derived results of this algorithm.

Additionally, $\Phi = e^{-3r_1 -3r_2 - 3r_3}$ is converted in line 12 as 
\begin{equation}
\begin{split}
\left(
\begin{aligned}
& 0,0,0,0,0,0 \\
& 3,3,3,0,0,0
\end{aligned}
\right)
\rightarrow\ & 0 \left(
\begin{aligned}
& 0,0,0,0,0,0 \\
& 3,3,3,0,0,0
\end{aligned}
\right)
\end{split}
\label{cf_1s3}
\end{equation}
The zero prefactor at the RHS of \eqref{cf_1s3} suggests this term will be excluded.

\subsection{Generating Slater complement functions from expanding $1-e^{-\gamma r}$ type $g$ function.}
\label{gen_slater_function}
The order of the complement functions could affect the result of overlap matrix based selection and subsequently the electronic energy.  We present the computational routes in Algorithms \ref{alg_gen_slater}, \ref{alg_gen_gauss}, and \ref{alg_sel}.

In the previous works \cite{wang2025variational,wang2026free,wang2026variational}, the computations are within the hydrogen and helium ground states.

For a more-than-two-electron system, 
\begin{itemize}
\item [(i)] the different radial powers and exponents in the initial wavefunction 
\item [(ii)]     the effect of $1/r$ in the electron-nuclear attraction operator acting on the $n$s orbital ($n>1$) 
\item [(iii)]    series of antisymmetrization permutations
\end{itemize}
 lead to additional procedures of the workflow of the Gaussian-expanded free complement method.

The current algorithms are limited to s-type orbitals. Hence, no $l_i, m_i$  is presented in Algorithm \ref{alg_gen_slater}. The generalization to general angular momentum is expected to be straightforward, as outlined in Subsection \ref{outline}.

In Algorithm \ref{alg_gen_slater}, the input for the lithium ground state is $n=m=0$ in \eqref{psi_form_li}. Hence, the exponents and radial powers are $(\zeta_{1s}, \zeta_{1s}, \zeta_{2s})$ and $(0,0,1)$, respectively. 

In line 3 of Algorithm \ref{alg_gen_slater},  $\prec_g$ stands for the order of the tuples of powers of the $g$ functions, $(n_i, \cdots, n_{ij})$, not the order of $n_i, \cdots, n_{ij}$. For example, in the lithium ground state at $m=1$ with ascending $\prec_g$ is: ((0, 0, 0, 0, 0, 1), (0, 0, 0, 0, 1, 0), (0, 0, 0, 1, 0, 0), (0, 0, 1, 0, 0, 0), (0, 1, 0, 0, 0, 0), (1, 0, 0, 0, 0, 0)). In descending $\prec_g$,  it is: ((1, 0, 0, 0, 0, 0), (0, 1, 0, 0, 0, 0), (0, 0, 1, 0, 0, 0), (0, 0, 0, 1, 0, 0), (0, 0, 0, 0, 1, 0), (0, 0, 0, 0, 0, 1)). In either case, the consecutive integers correspond to $n_1, n_2, n_3, n_{12}, n_{13}, n_{23}$.

In line 13 of Algorithm \ref{alg_gen_slater},  the expansion order is from the \texttt{expand} function of the \texttt{SymPy} software for the previous works \cite{wang2025variational,wang2026free}. To increase the software independence of the present algorithm, we add \ref{alg_gen_slater} with ascend/descend options, besides the previous \texttt{SymPy} order.

In line 14 of Algorithm \ref{alg_gen_slater},  we denote the power and exponent with $^{\tilde{}}$ after expanding power series of $1-e^{-\gamma r}$.

Notice that in the input of Algorithm 1 of the previous work \cite{wang2025variational}, the spatial part of the complement function is ordered as $\Phi_0$, $g_{12} \Phi_0$, $(1+P_{12}) g_1 \Phi_0$. In terms of the powers of the $g$ functions $(0,0,0), (0,0,1), (1,0,0)$. This introduced different orders of the geminal and orbital powers. In the present work, ascending or descending $\prec_g$ will lead to $(0,0,0), (0,0,1), (0,1,0)$ or $(0,1,0), (0,0,1), (0,0,0)$. Nevertheless, comparing the numerical results in the previous work \cite[Tables 1 and 2]{wang2026free} with Tables \ref{tab1} and \ref{tab2} and Appendix of the present work, this difference does not change the overall picture of the hierarchically decontracted Gaussian expanded complement function method.

\begin{threeparttable}
\begin{algorithm}[H] 
    \caption{Generating Slater-product complement functions for $N$-electron system} \label{alg_gen_slater}
    \KwIn{ $\{  \zeta_i , \mu_i, \gamma_i, \gamma_{ij}, \Theta, n  \}  $ \tnote{a}   \\
    $\prec_g$,  $\prec_V$, $\prec_{\mathrm{Slater}}$
    }
    \KwOut{\texttt{Slater\_products}. }
    \texttt{complement\_specs} = () \;
    \For{$m = 1$ to $n$}{
     $( ( n_i, n_{ij}  ) )$  $\Leftarrow$ \,    $(\sum_i g_i + \sum_{i<j} g_{ij})^m$ as  $\prec_g$ \;
    \For{ $(  n_i, n_{ij} )$  in  $( ( n_i, n_{ij}  ) )$  }{      \texttt{complement\_specs} $\leftarrow$ $(  (n_i, n_{ij}) , (\gamma_i, \gamma_{ij}),  (\mu_i), (\zeta_i) )$\;   
     \For{$m' = 1$ to $m$}{
     $ ( ( \mu_i -\nu_i  ) ) \Leftarrow (\sum_j 1/r_j)^{m'}  \prod_i r_i^{\mu_i}   $  as $\prec_V$\;
  
    \For{  $( \mu_i -\nu_i  ) $ in   $( (\mu_i -\nu_i  ) ) $    }{   \texttt{complement\_specs} $\leftarrow$ $ (   (n_i, n_{ij}) , (\gamma_i, \gamma_{ij}),  (\mu_i-\nu_i), (\zeta_i) )    $ \;  } }
    }
    }
    Remove duplicated entries in \texttt{complement\_specs} by the algorithm in Subsection \ref{unique}\;
    Convert indices  in \texttt{complement\_specs} into symbolic form of Eq. \eqref{g_psi0}\;
    \texttt{Slater\_products} = ()\;
    \For{\texttt{\upshape  complement\_spec} in \texttt{\upshape  complement\_specs} }{
     $  \prod_{j<k}     e^{- n_{jk} \gamma_{jk}}  \prod_{i}  r_{i}^{\mu_{i} - \nu_{i}}   e^{- (n_{i} \gamma_{i} + \zeta_{i}) r_{i}}   $ as $\prec_{\mathrm{Slater}}$  $\Leftarrow$   \texttt{\upshape  complement\_spec}\;
    $(  (\tilde \mu_i, 0), (\tilde \zeta_i, \tilde \zeta_{jk} )   ) $   $ \Leftarrow   \prod_{j<k}     e^{- n_{jk} \gamma_{jk}}  \prod_{i}  r_{i}^{\mu_{i} - \nu_{i}}   e^{- (n_{i} \gamma_{i} + \zeta_{i}) r_{i}}       $ \; 
    \texttt{Slater\_products} $\leftarrow$  Remove duplicated entries in $(  (\tilde \mu_i, 0), (\tilde \zeta_i, \tilde \zeta_{jk} )   ) $  by the algorithm in Subsection    \ref{unique}\;
    }
    Remove duplicated duplicated entries in   \texttt{Slater\_products}  \;
    \Return  \texttt{Slater\_products} 
\end{algorithm}
\begin{tablenotes}
  \item[a] Besides nuclear charge, Gaussian expansion length. \\
\end{tablenotes}
\end{threeparttable}

\subsection{Generating Gaussian-expanded complement functions from Slater complement functions} \label{gen_gauss_function}

In Algorithm \ref{alg_gen_gauss}, we present the workflow of generating the Gaussian expanded complement functions in the present work. We have included the radial power of the Slater function comparing to the previous work \cite{wang2025variational}. The incorporation of angular momentum is expected to be straightforward.

In lines 5 and 6 of Algorithm \ref{alg_gen_gauss}, 1.0 $\times 10^{-9}$ numerical threshold for the relative difference, namely $|a-b|/\mathrm{max}(|a|,|b|)< 1\times 10^{-9}$  \cite{python_math} has been used to determine if the exponents are numerical the same for decontraction/contraction, as in the previous works \cite{wang2025variational,wang2026free}.

When using the geminal induced expansion length \eqref{inequal_ng}, the $n_{ij}>0$ condition is determined by slightly different ways in the power and pair approaches mentioned in Section \ref{result}. Namely, in the power approach, we calculate the sum of geminal exponents then divided by 0.5, round to integer to obtained the total power. In the pair approach, it is by comparing each geminal exponent, if it is close to 0.5 within 1.0 $\times 10^{-9}$ numerical threshold for the relative difference mentioned in the previous paragraph, the number of pairs is added by 1. The total power or number of pairs will be used to load the associated expansion length. The 1.0 $\times 10^{-9}$ numerical threshold for the total power of geminal exponents is not expected to lead to numerical difference.

In line 7 of Algorithm \ref{alg_gen_gauss}, $\times$ stands for the Cartesian product. 

\begin{threeparttable}
\begin{algorithm}[H] 
    \caption{Generating Gaussian complement functions for $N$-electron system} \label{alg_gen_gauss}
    \KwIn{ \texttt{Slater\_products}, \texttt{initial\_wavefunction}, $\prec_{\mathrm{STO-}n\G}$
    }
    \KwOut{\texttt{Formatted\_Gauss\_products}. }
    \texttt{Gauss\_products} = () \;
    \For{\texttt{\upshape Slater\_product} in \texttt{\upshape{Slater\_products}}  }{
      \texttt{stacks} = ()\;
      \mbox{\texttt{decontraction\_choice} $\Leftarrow$ \, \texttt{Slater\_product}, \texttt{initial\_wavefunction}\;}
    \For{ $m=1$  to  $ n_{\textrm{\upshape{orbital}}} + n_{\textrm{\upshape{geminal}}}$  }{       
    \texttt{stacks} $\leftarrow$ ($\alpha$) as $\prec_{\mathrm{STO-}n\G}$  $\Leftarrow$ ($\tilde \mu$, $\tilde \zeta$) $\in$ \texttt{Slater\_product}[$m$], \KwSty{if} \texttt{decontraction\_choice}[$m$] = True\;
    \texttt{stacks} $\leftarrow$ ($\alpha$, $c$) as $\prec_{\mathrm{STO-}n\G}$ $\Leftarrow$ ($\tilde \mu$, $\tilde \zeta$) $\in$ \texttt{Slater\_product}[$m$], \KwSty{if} \texttt{decontraction\_choice}[$m$] = False\;
    }
        \texttt{Gauss\_products}  $\leftarrow$ remove duplicated entries in \texttt{stacks}[1] $\times \cdots \times $\texttt{stacks}[$n_{\textrm{\upshape{orbital}}} + n_{\textrm{\upshape{geminal}}} $]\tnote{a} \; 
    }
    \texttt{Formatted\_Gauss\_products} $\Leftarrow$   \texttt{ Gauss\_products}\tnote{b} \;
    \Return  \texttt{Formatted\_Gauss\_products} 
\end{algorithm}
\begin{tablenotes}
  \item[a] First entry as the product of the first element of each \texttt{stacks}[i], then vary the right most entry, as the order of product function of \texttt{itertools} in \texttt{Python}. \\
  \item[b] \texttt{Formatted} means in the forms of (24) - (26) in Ref. \cite{wang2026free} . \\
\end{tablenotes}
\end{threeparttable}

\subsection{Selection of Gaussian expanded complement function based on overlap matrix and subsequent treatment of the secular equation} \label{overlap_sel_secular}

In Algorithm \ref{alg_sel}, we present the workflow of overlap matrix selection of the Gaussian expanded complement functions \cite{von2008trapped,von2009correlated,rakshit2012hyperspherical,mitroy2013theory,Kalaee2014,Mosegaard2018,moriya2023novel,coomar2022quantum,wang2025variational,wang2026free}. In the previous work, no absolute value of the overlap between two complement function was used in the numerator term \cite[line 9 of Algorithm 2]{wang2025variational}. This is not expected to produce any numerical issue in the previous work \cite{wang2026free}, since all linear combination coefficients of the 1s STO-$n$G orbital are found to be positive \cite{o1966gaussian,hehre1969self,fernandez1988accurate}. For 2s, they are not \cite{stewart1970small,fernandez1988accurate}. In the present work. We have added the absolute value as in Ref. \cite{bubin2013born}.

\begin{threeparttable}
\begin{algorithm}[H] 
    \caption{Selection based on overlap matrix for $N$-electron system} \label{alg_sel}
    \KwIn{  \texttt{Formatted\_Gauss\_products}
    }
    \KwOut{\texttt{Selected\_Formatted\_Gauss\_products}. }
    \texttt{Duplicated\_indices} = () \;
    \texttt{Formatted\_Gauss\_products} = () \;
    \For{ $I=1$ to $M^{\mathrm{before}}$\tnote{,a}  }{
      \If{$I\in $\texttt{ \upshape{Duplicated\_indices}} }{ continue\;}  
    \For{ $J=I+1$  to  $M^{\mathrm{before}}$ }{    
       \If{$J\in $\texttt{ \upshape{Duplicated\_indices}} }{ continue\;}  
    \If{ $\frac{ |\braket{ \Phi_I | \mathcal{A} | \Phi_J}|}{\braket{ \Phi_I | \mathcal{A} | \Phi_I}^{1/2} \braket{ \Phi_J | \mathcal{A} | \Phi_J}^{1/2}\tnote{,b}}$ > \upshape{ overlap\_threshold \cite{bubin2013born}} }{ \texttt{Duplicated\_indices}   $\leftarrow$ $J$\;}
    }
    }
    \For{ $I=1$ to $M^{\mathrm{before}}$  }{
      \If{$I\in $\texttt{ \upshape{Duplicated\_indices}} }{ continue\;}  \texttt{Selected\_Formatted\_Gauss\_products}  $\leftarrow$ \texttt{Formatted\_Gauss\_products}[$I$]\; }
    \Return  \texttt{Selected\_Formatted\_Gauss\_products} 
\end{algorithm}
\begin{tablenotes}
  \item[a] Number of complement function of the input, \texttt{Formatted\_Gauss\_products}. \\
  \item[b] $\braket{ \Phi_I | \mathcal{A} | \Phi_I}^{1/2} , \forall I$ are computed by a single loop before line 3 to reduce the computational cost. \\
\end{tablenotes}
\end{threeparttable}

In the numerical calculations of the overlap matrix based selection and energy calculation via the secular equation, we do not include the prefactor $N!$ in Eq. \eqref{one_fold_permu}.  After computing the Hamiltonian and overlap matrices via Eq. \eqref{one_fold_permu},  we perform the substitutions
\begin{align}
  H_{ij} &\leftarrow H_{ij}  / \sqrt{S_{ii} S_{jj}} \\
  S_{ij} &\leftarrow S_{ij}  / \sqrt{S_{ii} S_{jj}} 
\end{align}
to scale the matrix elements as normalized complement functions. These are used in the previous works \cite{wang2025variational,wang2026free}.

\subsection{Symmetries of integrals}
In the previous work \cite{wang2026free}, we noticed for the helium ground state, the overlap, the sum of electron-nucleus attraction, and electron-electron repulsion integrals have 16-fold symmetries; the sum of the kinetic integrals has 4-fold symmetries.

In general, consider
\begin{align}
    I_{\substack{a_1 \cdots a_N a_{12} \cdots a_{N-1 N} \\
    b_1\cdots b_N    b_{12} \cdots b_{N-1N} } } := \braket{ e^{- \sum_{1 \leq i \leq N} a_i r_i^2 - \sum_{1 \leq i<j \leq N }  a_{ij} r_{ij}^2 } | \hat O |  e^{- \sum_{1 \leq i \leq N} b_i r_i^2 - \sum_{1 \leq i<j \leq N }  b_{ij} r_{ij}^2 } } \label{int_ne}
\end{align}
no permutation nor antisymmetrization inside the ket or bra vector.

For the overlap, the electronic sum of electron-nucleus attractions, the sum of electron-electron repulsions, $\hat O = \hat 1, \sum_{1 \leq i \leq N} \hat V_{\en,i}, \sum_{ 1 \leq i<j \leq N } \hat V_{\ee,ij}$, the integral \eqref{int_ne} is invariant under exchanging $a_i \leftrightarrow b_i$, $a_{ij} \leftrightarrow b_{ij}$ and permutation $\left[\begin{smallmatrix} 1& \cdots &N\\ \sigma_1 & \cdots & \sigma_N \end{smallmatrix}\right]$.  This leads to $2^{N + N(N-1)/2} N! = 2^{N(N+1)/2} N!$ fold symmetries. The aforementioned exchange symmetries cover the bra-ket Hermitian symmetry. Notice the electronic sum of the electron-nucleus attraction operator $\sum_i \hat V_{\en,i}$, whether sum over the nucleus does not affect this type of symmetry. In addition, the symmetries in the three-electron integral via renaming variables have been discussed in Ref. \cite{puchalski2010applications}.

For the sum of the one-electron kinetic operators, $\sum_{1 \leq i \leq N} \hat T_i$,  the bra-ket Hermitian  and permutation invariance lead to $2N!$ fold symmetries. In addition, consider the incomplete sum, i.e., $\sum_{1 \leq i \leq k <N} \hat T_i$. Outside this summation of kinetic operators, there is $2^{(N-k)(N-k+1)/2} k!$ fold symmetry which resembles $\hat O = \hat 1$.  By further adopting with $2k!$ fold symmetries of this partial summed kinetic operators and $C(N,k)$ combination of the partial summation, this leads to $2 N! 2^{(N-k)(N-k+1)/2},  1 \leq k \leq N$ fold symmetries. This expression includes $k=N$, since it is also valid.

Furthermore, with the help from Claude Opus 5.5 \cite{claude}, the overlap, electronic sum of electron-nucleus attractions, and the sum of electron-electron repulsions have a semidirect product structure $\mathbb Z_2^{N(N+1)/2} \rtimes S_N$. Here $\mathbb Z_2$ is the permutation of ket and bra variables associated to a given coordinate, e.g., $a_i \leftrightarrow b_i $. $S_N$ is the symmetric group at order of the number of electrons. The group element operates on the target slots (integrated variables).

For example, given $(a_1, a_2, a_{12}, b_1, b_2, b_{12})$ for the helium ground state. $P$(slot\_1, slot\_2)$^{-1}$ gives $(a_2, a_1, a_{12}, b_2, b_1, b_{12})$. $\mathbb Z_2$(first slots on ket/bra)  $P$(slot\_1, slot\_2)$^{-1}$ gives  $(b_2, a_1, a_{12}, a_2, b_1, b_{12})$. $P$(slot\_1, slot\_2)  $\mathbb Z_2$(first slots on ket/bra)  $P$(slot\_1, slot\_2)$^{-1}$ gives $(a_1, b_2, a_{12}, b_1, a_2, b_{12})$. This can be obtained by $\mathbb Z_2$(second slots on ket/bra). This supports $\mathbb Z_2^3$ for the helium ground state is a normal subgroup.

For $P$(slot\_1, slot\_2) $\in S_2$, $\mathbb Z_2^{-1}$(first slots on ket/bra) on  $(a_1, a_2, a_{12}, b_1, b_2, b_{12})$ gives $(b_1, a_2, a_{12}, a_1, b_2, b_{12})$. $P$(slot\_1, slot\_2)  $\mathbb Z_2^{-1}$(first slots on ket/bra) gives  $(a_2, b_1, a_{12}, b_2, a_1, b_{12})$.     $\mathbb Z_2$(first slots on ket/bra)  $P$(slot\_1, slot\_2)  $\mathbb Z_2^{-1}$(first slots on ket/bra) gives $(b_2, b_1, a_{12}, a_2, a_1, b_{12})$. This cannot be obtained by any element in $S_2$ on $(a_1, a_2, a_{12}, b_1, b_2, b_{12})$. Thus, $S_2$ is not a normal subgroup in this case.

In the present work, we have used the formulae in Ref. \cite{mitroy2013theory} for the integral computations. Explicit expressions for $3\times 3 $ matrix, including determinant, inverse with adjugate, multiplication, and trace,  are used to accelerate the calculations. The Cholesky decompositions are used in other dimensions, especially for the helium ground state in the present work. 

Preliminary advances have been made for including the aforementioned symmetries for the overlap integrals in the overlap matrix based selection. For the subsequent evaluations of the integrals in the secular equation, $\mathbf{HC} = \mathbf{SCE}$, only the Hermitian symmetry was used by computing the upper triangular parts of the overlap and Hamiltonian matrices (let the remaining matrix elements equal to the Hermitian parts).  We plan to publish the scheme of symmetry adaptations with the integral evaluations after better incorporating these symmetries in future \cite{wang_unpublished}.  The possible numerical difference via symmetry adaptations is expected to be insignificant.

\section{Implementations} \label{implementation}
The programming software \texttt{Python} version 3.12.3  \cite{van1991interactively} is used. \texttt{SymPy} version 1.14.0  \cite{10.7717/peerj-cs.103} is  used in the symbolic computations of  $1 - e^{-r}$ type $g$ functions. \texttt{Mpmath} multi-precision library version 1.3.0 \cite{mpmath} is used in generating the complement functions with 50 digit precision. \texttt{NumPy} version 1.26.2 \cite{harris2020array} and \texttt{Numba} version 0.61.2 \cite{10.1145/2833157.2833162} are used for double precision computations and enhancing the efficiency respectively. The Intel Math Kernel Library (\texttt{MKL}) with the single dynamic library (\texttt{SDL}) version 2023.1 \cite{intel_mkl, intel_sdl} was used as the dependency of \texttt{NumPy} in the canonical orthogonalization approach of solving the secular equation. The integrals evaluations via the \texttt{njit} function of \texttt{Numba} and \texttt{MKL-SDL} usage are paralleled with 12 cores. 

The software \texttt{Cursor} combined with large language models \cite{cursor} has been used in generating the \texttt{Python} code.

\section{Numerical results and discussions} \label{result}

In this Section, the energies (except the concept subchemical accuracy, 0.1 kcal mol$^{-1}$) \cite{koput1995ab}) and the minimum value of the overlap matrix, $s_\mathrm{min}$, will be rounded to even as in the present work  \cite{python_format,python_round,goldberg1991every}. In the present work, $1\times 10^{-12}$ threshold of throwing away basis functions in the canonical orthogonalization is adopted by default, unless explicitly mentioned other value. For the STO-$n$G expansion, we have used the results from Ref. \cite{stewart1970small} for $n=6$, since the $s$ and $p$ orbitals are separately optimized. For $n=8, 10, 12, 14$, the results from Ref. \cite{fernandez1988accurate} are used, since Ref. \cite{o1966gaussian} has $n=8$ and $n=10$ but only used single precision.

To better indicate the influence of geminal induced expansion length and considering the size of table, we use the error with respect to the reference values \cite{schwartz2006further,nakashima2007solving,kurokawa2008solving,nakashima2008accurately,doi:10.26434/chemrxiv.15010182/v3,wang2012variational} than the total energy. For example, if two errors $1.10 \times 10^{-3}$ and $1.20 \times 10^{-3}$, we can infer from inspection that the relative difference between these two errors is roughly about $10 \%$. If total energies are used, this ratio would possibly require some subtraction and less obvious from inspection.

To this end, we used the reference energies via helium \cite{schwartz2006further,nakashima2007solving,kurokawa2008solving,nakashima2008accurately,doi:10.26434/chemrxiv.15010182/v3} and lithium \cite{wang2012variational}, $-2.90372437703$ and $-7.478 060 323 91$, respectively.

The inequality \eqref{inequal_ng} contains exponential possibilities with respect to the number of coordinate variables for each complement function in the form of product of Slater functions. For example, if we denote the spatial part of the complement function as $e^{-n_1 \zeta_1 r_1-n_2 \zeta_2 r_2-n_{12} \zeta_{12} r_{12} }  \Phi_0  $ as $(n_1,n_2,n_{12})$. For (0,0,0), $n_{\G_1}$ = 3,4,5; $n_{\G_2}$ = 3,4,5; $n_{\G_{12}}$ = 3,4,5, there is $3^3=27$ combinations, excluding equivalent terms by the antisymmetrization. For 10 complement functions with such choices of expansion lengths, there are about 270 possibilities.

Thus, we restrict ourselves to a preliminary form in the present scope. We consider the following form of the expansion lengths with the presence of the geminal function: $(n_{\G_1}^{(1)}, \cdots, n_{\G_N}^{(1)} ) $, 
$(n_{\G_1}^{(2)},  \cdots, n_{\G_N}^{(2)} ) $, $\cdots$, $(n_{\G_{12}}^{(1)},  \cdots, n_{\G_{N-1, N}}^{(1)} )$ , $(n_{\G_{12}}^{(2)},  \cdots, n_{\G_{N-1, N}}^{(2)} )$, $\cdots$.  The superscripts $(1), (2),\cdots$  are level parameters. They correspond to
\begin{itemize}
\item [(power)] the sum of powers of the geminal function in the Slater product form complement function. For example, $(2)$ for $n_{12}=n_{13}=1, n_{23}=0$ and  $n_{12}=2, n_{13}=n_{23}=0$ in 

$e^{-n_1 \zeta_1 r_1-n_2 \zeta_2 r_2 -n_3 \zeta_3 r_3 -n_{12} \zeta_{12} r_{12} -n_{13} \zeta_{13} r_{13} -n_{23} \zeta_{23} r_{23}   }  \Phi_0  $, this corresponds to the order $n$ of the FC expansion; or
\item [(pair)] the number of pairs of geminal functions.  For example,  $(2)$ 

for $e^{-n_1 \zeta_1 r_1-n_2 \zeta_2 r_2 -n_3 \zeta_3 r_3 -n_{12} \zeta_{12} r_{12} -n_{13} \zeta_{13} r_{13} -n_{23} \zeta_{23} r_{23}   }  \Phi_0  $ when 

$n_{12} >0, n_{13}>0, n_{23} = 0$ 
\end{itemize}

We shall use $n_\G^{\orb}$ and $n_\G^{\gem}$ to denote $(n_{\G_1}^{(1)}, \cdots, n_{\G_N}^{(1)} ) $,$(n_{\G_1}^{(2)}, \cdots, n_{\G_N}^{(2)} ) $, $\cdots$ and $(n_{\G_{12}}^{(1)},  \cdots, n_{\G_{N-1, N}}^{(1)} )$ , $(n_{\G_{12}}^{(2)},  \cdots, n_{\G_{N-1, N}}^{(2)} )$, $\cdots$, respectively. For limiting case all geminal induced orbital and geminal expansion lengths equal within each other, cases (power) and (pair) are the same. We denote $ n_{\G_1}^{(1)} =  \cdots = n_{\G_N}^{(1)} = n_{\G_1}^{(2)} = \cdots = n_{\G_N}^{(2)} = \cdots = n_\G^{\orb} $ or  $ n_{\G_{12}}^{(1)} =  \cdots = n_{\G_{N-1, N}}^{(1)} = n_{\G_{12}}^{(2)} = \cdots = n_{\G_{N-1,N}}^{(2)} = \cdots = n_\G^{\gem} $.

\subsection{Helium ground state} \label{he}

The initial wavefunction $\psi_0$ via Eq. \eqref{psi0_general} is
\begin{align}
   \psi_0 &= \mathcal A \left[ e^{-\zeta (r_1 + r_2)   } \Theta  \right] \label{psi0_he} \\
  \Theta   &=   \chi_+ (1) \chi_-(2)  - \chi_- (1) \chi_+(2) 
\end{align}
where  $\zeta  = 1.6875$ \cite{nakatsuji2024exact} is adopted. For the $1-e^{-\gamma r}$ type $g$ function, $\gamma_{i} = 2 - 1.6875 = 0.3125$ and $\gamma_{ij} = 0.5$ \cite{nakatsuji2024exact}  are adopted.

In this Subsection, we only consider all expansion lengths with the presence of geminal function equal. Namely, $n_{\G_1}^{(1)} = n_{\G_2}^{(1)} = n_{\G_{12}}^{(1)} = n_{\G_{1}}^{(2)} =  n_{\G_{2}}^{(2)} = n_{\G_{12}}^{(2)} = \cdots $. The numerical results are presented in Tables \ref{tab1} and \ref{tab2}.

According to the energies in Tables \ref{tab1} and \ref{tab2} suggest, for order $n=1$, using lower expansion lengths of yields insignificant difference of energies. For example, up to about 2\% with respect to the energy of $n_\G = n_\G^{\orb} = n_\G^{\gem}$ in 0.95 overlap selection threshold and $n=10$.

For $n=2,3$, lowering $n_\G^{\orb}$ and $n_\G^{\gem}$ leads to larger difference than $n=1$. Notice for $n=2$, $n_\G = n_\G^{\orb} = n_\G^{\gem} = 10$ with 0.95 threshold of overlap matrix selection, $\Delta E$ is at the fringe below the subchemical accuracy (0.1 kcal mol$^{-1}$) \cite{koput1995ab}. This makes further lowering $n_\G^{\orb}$ and $n_\G^{\gem}$ easily exceed the subchemical accuracy on $\Delta E$.

Notice, occasionally the energy from geminal induced expansion lengths can be more accurate than without introducing this variation. For example, $n=3$ and $n_\G=10$, $n_\G^{\orb} = n_\G^{\gem}$ from 10 to 8, in 0.95 overlap selection threshold,  $\Delta E$ decreases from $2.2 \times 10^{-5}$ to $1.5 \times 10^{-5}$ in Table \ref{tab1}. Increasing the overlap matrix selection thresholds to 0.98, 0.99, and 0.995 in Tables \ref{tab1} and \ref{tab2}, within this $n_\G, n_\G^{\orb}, n_\G^{\gem}  $, it does not happen.

Furthermore, for $n = 3$, $n_\G = 14$, $n_\G^{\orb} = n_\G^{\gem}$ from 14 to 12 and further to 10, in 0.995 overlap matrix selection threshold in Table \ref{tab2}, $s_{\min}$ decreases noticeably. 

According to Appendix of the present work, different ordering could vary the energy and $s_{\min}$. For the question how different orderings would further affect the behaviors of varying $n_\G^{\orb}$ and $n_\G^{\gem}$, it may be more theoretically motivated to 
looking into other form of overlap matrix based selection. Namely, keeping the selected Gaussian expanded complement functions at low order of $n$, and compare each new complement function at higher order of $n$. This is closer to the original usage of this method \cite{von2008trapped,von2009correlated,rakshit2012hyperspherical,mitroy2013theory,Kalaee2014,Mosegaard2018,moriya2023novel,coomar2022quantum}.

Notice in Table \ref{tab2}, some values of $s_{\min}$ are negative in the present double precision computations. Nevertheless, after removal the functions below the numerical threshold by the canonical orthogonalization method, the results still appear normal. The normal is justified by 
\begin{itemize}
\item[(i)] the energy associated with negative $s_{\min}$ is similar to the energy with positive $s_{\min}$ in the same $n$, $n_{\G}$ and overlap matrix selection threshold. For example, 0.99 overlap selection threshold, $n=2$, $n_\G=14$,  $ n_\G^\orb =  n_\G^\gem = 10$ with $s_{\min} = -4.6 \times 10^{-15}$, the energy does not have drastic difference to  $ n_\G^\orb =  n_\G^\gem = 12$ or $ n_\G^\orb =  n_\G^\gem = 14$, within $n=2$, $n_\G=14$, and 0.99  overlap selection threshold;
\item[(ii)] the energy associated with negative $s_{\min}$ is similar to the previous work with 50 digit precision \cite{wang2026free}. Though the previous work used a different ordering of complement functions as discussed in Subsection \ref{gen_slater_function}, the energy difference, according to Appendix of the present work, is insignificant for the purpose of evaluating the result of removal basis function of canonical orthogonalization. 
\end{itemize}

\begin{table}[H]
\centering
\setlength{\tabcolsep}{3pt}
\caption{Results of the variational calculations via the Gaussian expanded complement functions, hierarchical decontraction, and geminal induced expansion length for the ground state of the helium atom. 0.95 and 0.98 are the thresholds for the overlap selections in Algorithm \ref{alg_sel}.  Orderings $\prec_g = \prec_V = \prec_{\mathrm{Slater}} = \prec_{\mathrm{STO-}n\G} = \uparrow$ are adopted. $n_{\G}$, $n_G^{\orb}$, and $n_\G^{\gem}$ stand for the expansion length without the geminal function, the expansion length for orbital Slater function with the presence of geminal function, and the expansion length for the geminal Slater function with presence of geminal function. $M^\before$ and $M^\after$ are denoted to the number of the complement functions before and after the overlap selection, respectively.  $s_{\mathrm{min}}$ stands for the minimum eigenvalue of the overlap matrix after the overlap selection. $\Delta E$ stands for the computed electronic energy \cite[p. 43]{szabo1996modern} subtracts the reference energy. The values of $s_{\mathrm{min}}$  and $\Delta E$ are rounded to even \cite{python_format,python_round,goldberg1991every}.  The integer in square bracket is denoted to 10 to the power of the given integer \cite{safronova2011correlation}.   }

\begin{threeparttable}
\begin{tabular}{ccccc ccc ccc}
\hline
\multicolumn{5}{c}{} & \multicolumn{3}{c}{0.95} & \multicolumn{3}{c}{0.98} \\
$n$ & $n_\G$  & $n_\G^{\orb}$  &  $n_\G^{\gem}$   & $M^\before $ & $M^\after$  &  $s_{\mathrm{min}}$ & $\Delta E$ & $M^\after $  & $s_{\mathrm{min}}$  &  $\Delta  E$   \\ \hline
1 & 10 & 10 & 10 & 21 & 13 & $2.7[-3]$ & $5.27[-3]$ & 15 & $2.5[-4]$ & $4.36[-3]$ \\
 &  & 8 & 8 & 19 & 12 & $3.7[-3]$ & $5.21[-3]$ & 14 & $2.6[-4]$ & $4.40[-3]$ \\
 &  & 6 & 6 & 17 & 11 & $4.6[-3]$ & $5.40[-3]$ & 13 & $1.0[-4]$ & $4.60[-3]$ \\
 & 14 & 14 & 14 & 29 & 18 & $1.0[-3]$ & $4.66[-3]$ & 21 & $1.6[-4]$ & $4.32[-3]$ \\
 &  & 12 & 12 & 27 & 17 & $1.0[-3]$ & $4.65[-3]$ & 20 & $1.6[-4]$ & $4.32[-3]$ \\
 &  & 10 & 10 & 25 & 16 & $1.0[-3]$ & $4.76[-3]$ & 19 & $1.6[-4]$ & $4.31[-3]$ \\
2 & 10 & 10 & 10 & 196 & 71 & $1.7[-5]$ & $1.53[-4]$ & 98 & $1.6[-7]$ & $1.01[-4]$ \\
 &  & 8 & 8 & 156 & 69 & $8.1[-6]$ & $1.70[-4]$ & 92 & $1.5[-7]$ & $9.82[-5]$ \\
 &  & 6 & 6 & 124 & 62 & $1.2[-5]$ & $2.79[-4]$ & 78 & $3.0[-8]$ & $1.64[-4]$ \\
 & 14 & 14 & 14 & 358 & 126 & $4.7[-6]$ & $9.99[-5]$ & 174 & $9.9[-9]$ & $2.22[-5]$ \\
 &  & 12 & 12 & 302 & 125 & $5.4[-6]$ & $7.03[-5]$ & 170 & $2.9[-9]$ & $2.28[-5]$ \\
 &  & 10 & 10 & 254 & 112 & $1.5[-6]$ & $1.25[-4]$ & 158 & $5.5[-10]$ & $3.61[-5]$ \\
3 & 10 & 10 & 10 & 1066 & 102 & $1.2[-5]$ & $5.98[-5]$ & 195 & $2.5[-8]$ & $2.20[-5]$ \\
 &  & 8 & 8 & 690 & 93 & $4.5[-6]$ & $8.13[-5]$ & 164 & $2.1[-8]$ & $1.51[-5]$ \\
 &  & 6 & 6 & 438 & 81 & $4.8[-6]$ & $6.49[-5]$ & 126 & $1.9[-8]$ & $4.04[-5]$ \\
 & 14 & 14 & 14 & 2444 & 181 & $1.7[-6]$ & $1.94[-5]$ & 342 & $7.5[-10]$ & $2.83[-6]$ \\
 &  & 12 & 12 & 1748 & 174 & $3.3[-6]$ & $1.98[-5]$ & 322 & $1.0[-9]$ & $1.71[-6]$ \\
 &  & 10 & 10 & 1224 & 142 & $1.3[-6]$ & $3.19[-5]$ & 268 & $5.0[-10]$ & $3.00[-6]$ \\
\hline
\end{tabular}
\label{tab1}
\end{threeparttable}
\end{table}

\begin{table}[H]
\centering
\setlength{\tabcolsep}{3pt}
\caption{
Results of the variational calculations via the Gaussian expanded complement functions, hierarchical decontraction, and geminal induced expansion length for the ground state of the helium atom. 0.99 and 0.995 are the thresholds for the overlap selections in Algorithm \ref{alg_sel}.  Orderings $\prec_g = \prec_V = \prec_{\mathrm{Slater}} = \prec_{\mathrm{STO-}n\G} = \uparrow$ are adopted. $n_{\G}$, $n_G^{\orb}$, and $n_\G^{\gem}$ stand for the expansion length without the geminal function, the expansion length for orbital Slater function with the presence of geminal function, and the expansion length for the geminal Slater function with presence of geminal function. $M^\before$ and $M^\after$ are denoted to the number of the complement functions before and after the overlap selection, respectively.  $s_{\mathrm{min}}$ stands for the minimum eigenvalue of the overlap matrix after the overlap selection. $\Delta E$ stands for the computed electronic energy \cite[p. 43]{szabo1996modern} subtracts the reference energy. The values of $s_{\mathrm{min}}$  and $\Delta E$ are rounded to even \cite{python_format,python_round,goldberg1991every}. The integer in square bracket is denoted to 10 to the power of the given integer \cite{safronova2011correlation}.  
}
\begin{threeparttable}
\begin{tabular}{ccccccccccc}
\hline
\multicolumn{5}{c}{} & \multicolumn{3}{c}{0.99} & \multicolumn{3}{c}{0.995} \\
$n$ & $n_\G$  & $n_\G^{\orb}$  &  $n_\G^{\gem}$   & $M^{\before}$ & $M^\after$  &  $s_{\mathrm{min}}$ & $\Delta E$ & $M^\after$  & $s_{\mathrm{min}}$  &  $\Delta  E$   \\ \hline
1 & 10 & 10 & 10 & 21 & 17 & $2.5[-7]$ & $4.37[-3]$ & 18 & $2.4[-7]$ & $4.34[-3]$ \\
 &  & 8 & 8 & 19 & 16 & $2.5[-7]$ & $4.39[-3]$ & 17 & $2.4[-7]$ & $4.37[-3]$ \\
 &  & 6 & 6 & 17 & 15 & $2.5[-7]$ & $4.60[-3]$ & 15 & $2.5[-7]$ & $4.60[-3]$ \\
 & 14 & 14 & 14 & 29 & 24 & $9.2[-9]$ & $4.28[-3]$ & 25 & $9.2[-9]$ & $4.26[-3]$ \\
 &  & 12 & 12 & 27 & 23 & $9.2[-9]$ & $4.28[-3]$ & 23 & $9.2[-9]$ & $4.27[-3]$ \\
 &  & 10 & 10 & 25 & 21 & $9.2[-9]$ & $4.32[-3]$ & 22 & $9.2[-9]$ & $4.27[-3]$ \\
2 & 10 & 10 & 10 & 196 & 135 & $2.0[-13]$\tnote{a} & $2.72[-5]$ & 151 & $1.9[-13]$\tnote{b} & $2.50[-5]$ \\
 &  & 8 & 8 & 156 & 118 & $2.0[-10]$ & $3.33[-5]$ & 131 & $1.8[-13]$\tnote{c} & $3.02[-5]$ \\
 &  & 6 & 6 & 124 & 98 & $1.8[-13]$\tnote{d} & $6.37[-5]$ & 105 & $1.8[-13]$\tnote{e} & $5.79[-5]$ \\
 & 14 & 14 & 14 & 358 & 233 & $1.5[-14]$\tnote{f} & $1.25[-5]$ & 268 & $-2.5[-15]$\tnote{g} & $1.08[-5]$ \\
 &  & 12 & 12 & 302 & 217 & $3.2[-14]$\tnote{h} & $1.21[-5]$ & 245 & $-4.8[-15]$\tnote{i} & $1.15[-5]$ \\
 &  & 10 & 10 & 254 & 196 & $3.2[-15]$\tnote{j} & $1.27[-5]$ & 213 & $-4.1[-15]$\tnote{k} & $1.22[-5]$ \\
3 & 10 & 10 & 10 & 1066 & 367 & $1.8[-13]$\tnote{l} & $7.57[-6]$ & 497 & $8.3[-15]$\tnote{m} & $6.36[-6]$ \\
 &  & 8 & 8 & 690 & 323 & $2.7[-11]$ & $8.00[-6]$ & 430 & $1.2[-13]$\tnote{n} & $7.77[-6]$ \\
 &  & 6 & 6 & 438 & 244 & $1.8[-13]$\tnote{o} & $1.69[-5]$ & 304 & $9.1[-14]$\tnote{p} & $1.59[-5]$ \\
 & 14 & 14 & 14 & 2444 & 660 & $-1.5[-15]$\tnote{q} & $3.84[-7]$ & 882 & $4.3[-15]$\tnote{r} & $2.94[-7]$ \\
 &  & 12 & 12 & 1748 & 626 & $1.7[-15]$\tnote{s} & $3.58[-7]$ & 851 & $-2.1[-15]$\tnote{t} & $3.27[-7]$ \\
 &  & 10 & 10 & 1224 & 533 & $-3.7[-15]$\tnote{u} & $6.02[-7]$ & 687 & $-8.8[-16]$\tnote{v} & $5.68[-7]$ \\
\hline
\end{tabular}
\begin{tablenotes}
\item[a-v] The number of complement functions are throwing away by the canonical orthogonalization are 1, 1, 1, 1, 1, 1, 3, 1, 3, 2, 2, 1, 8, 4, 1, 2, 2, 28, 2, 30, 2, and 18, for the superscripts a, b, c, d, e, f, g, h, i, j, k, l, m, n, o, p, q, r, s, t, u, and v, respectively.
\end{tablenotes}
\end{threeparttable} \label{tab2}
\end{table}

\subsection{Lithium ground state}

The initial wavefunction $\psi_0$ via Eq. \eqref{psi0_general}  is
\begin{align}
   \psi_0 &= \mathcal A \left[ e^{-\zeta_1 (r_1 + r_2)  - \zeta_3 r_3 } \Theta  \right] \label{psi0_li_calc} \\
  \Theta   &=  \left[ \chi_+ (1) \chi_-(2)  - \chi_- (1) \chi_+(2) \right] \chi_+ (3)
\end{align}
where  $\zeta_1 =  \zeta_2 = 2.68$ and $\zeta_3 = 0.635$ \cite{PhysRevA.92.012512} are adopted.  For the $1-e^{-\gamma r}$ type $g$ function, $\gamma_1 = \gamma_2 = 3 - 2.68 = 0.32, \gamma_3 = 3- 0.635 = 2.365,   \gamma_{12} = \gamma_{13} = \gamma_{23} = 0.5$ \cite{nakatsuji2024exact}  are adopted. 

\subsubsection{Effect of 2s$\rightarrow$1s channel}

Notice for the initial wavefunction that includes 2s orbital, the $-Z/r$ term in the potential operator of the Hamiltonian can generate additional 1s orbital. This term resembles the $(a+br) e^{-c r}$ form of 2s orbital used in lithium computations \cite{PhysRev.36.878,PhysRev.42.33,wilson1933wave,guillemin1930einfache,PhysRev.70.197,tambasco1997variational}. 

To some extent, the linear combination of $\Phi_0$ and $1/r_3 \Phi_0$ resembles the nodal structure of 2s orbital caused by the orthogonal requirement of an one-particle Hermitian operator \cite{kellogg1916oscillations,alan1967,goddard1968new}. It has been found such node exists in natural orbital \cite{larsson1972natural}. Further studies see Ref. \cite{rasch2015fixed}.

Hence, we can have three possibilities:
\begin{itemize}
    \item [(i)] including 2s$\rightarrow$1s complement functions, with the decontraction and contraction choices \eqref{choice_decontraction} and \eqref{choice_contraction} covering both radial powers and exponents;
    \item [(ii)] including 2s$\rightarrow$1s complement functions, with decontraction and contraction choices modified \eqref{choice_decontraction} and \eqref{choice_contraction} that covers only exponents;
    \item [(iii)] excluding 2s$\rightarrow$1s complement functions, in this case, no difference for the aforementioned (i) and (ii). 
\end{itemize}

All these three approaches are expected to converge to the exact eigen energy when $n \rightarrow \infty$ and $n_\G \rightarrow \infty$. Since it is expected that the exponents cover $(0, \infty)$ with $n_\G \rightarrow \infty$, at high order, all exponents are decontracted thus allow all linear combinations, thus fulfills the sufficient condition of basis set completeness \cite{klahn1977convergence}.

Conceptually, the hierarchically decontraction may be viewed as corrections of a zero-order state. The aforementioned approach (ii) resembles a nodal reference wavefunction up to linear combination of contracted Gaussian expanded complement functions. Approach (i) is a relaxation of the Gaussian exponents via 2s$\rightarrow$1s. Approach (iii) would introduce nodal structure by further combination of Gaussian-expanded complement functions of various exponents \cite{goddard1968new}.

In Table \ref{tab3}, we present a comparison among three approaches for the lithium ground state with the initial wavefunction Eq. \eqref{psi0_li_calc}, $1-e^{-\gamma r}$ type $g$ function, and $p$-alone approach.

\begin{table}[H]
\centering
\setlength{\tabcolsep}{3pt}
\caption{
Results of the variational calculations via the Gaussian expanded complement functions, hierarchical decontraction, and geminal induced expansion length for the ground state of the lithium atom. Orderings $\prec_g = \prec_V = \prec_{\mathrm{Slater}} = \prec_{\mathrm{STO-}n\G} = \uparrow$ are adopted. $n_\G = n_\G^{\orb} = n_\G^{\gem} = 10$ is used. 0.95 is the default threshold for the overlap selections in Algorithm \ref{alg_sel}, except the last row. $M^\before$ and $M^\after$ are denoted to the number of the complement functions before and after the overlap selection, respectively.  $s_{\mathrm{min}}$ stands for the minimum eigenvalue of the overlap matrix after the overlap selection. $\Delta E$ stands for the electronic energy \cite[p. 43]{szabo1996modern}. The values of $E$ and $s_{\mathrm{min}}$ are rounded to even \cite{python_format,python_round,goldberg1991every}. The integer in square bracket is denoted to 10 to the power of the given integer \cite{safronova2011correlation}.  
}

\makebox[\linewidth][c]{
\begin{threeparttable}
\begin{tabular}{c cccc cccc cccc}
\hline
\multicolumn{1}{c}{} & \multicolumn{4}{c}{2s $\rightarrow$1s    } & \multicolumn{4}{c}{ 2s $\rightarrow$1s }  & \multicolumn{4}{c}{ 2s}  \\
\multicolumn{1}{c}{} & \multicolumn{4}{c}{ $\nu, \zeta$  } & \multicolumn{4}{c}{  $\zeta$ }  & \multicolumn{4}{c}{ }  \\
$n$ & $M^\before$  & $ M^\after$ & $s_{\mathrm{min}}$   &   $\Delta  E$   & $M^\before$    & $M^\after$   & $s_{\mathrm{min}}$   & $\Delta  E$    &     $M^\before$   &  $M^\after$  & $s_{\mathrm{min}}$     &   $\Delta  E$  \\ \hline
0 & 1 & 1 & $1.0[\phantom{-}0]$ & $6.03[-2]$ & 1 & 1 & $1.0[\phantom{-}0]$ & $6.03[-2]$ & 1 & 1 & $1.0[\phantom{-}0]$ & $6.03[-2]$ \\
1 & 361 & 160 & $4.3[-8]$ & $3.87[-3]$ & 82 & 47 & $9.6[-6]$ & $4.41[-3]$ & 41 & 29 & $1.4[-3]$ & $4.75[-3]$ \\
2 & 6921 & 1172 & $3.9[-10]$ & $1.58[-4]$ & 1782 & 529 & $4.4[-10]$ & $2.89[-4]$ & 891 & 346 & $2.5[-6]$ & $6.93[-4]$ \\
3 & 85221 & 3748 & $7.1[-12]$ & $8.72[-5]$ & 24912 & 2668 & $5.8[-12]$ & $8.50[-5]$ & 12731 & 1957 & $1.7[-8]$ & $2.80[-4]$ \\
 &  &  &  &  &  &  &  &  &  & 3497\tnote{a} & $1.1[-12]$\tnote{a} & $1.36[-4]$\tnote{a} \\
\hline
\end{tabular} \label{tab3}
\begin{tablenotes}
\item[a] 0.98 threshold of overlap selection   
\end{tablenotes}
\end{threeparttable}
}
\end{table}

Due to the consideration of other overlap selection approach in Subsection \ref{he}, we limit our scope to results with the overlap matrix selection threshold 0.95 in Table \ref{tab3} (besides one comparison with 0.98).  According to the data in Table \ref{tab3}, both approaches (i) and (ii) can reach the subchemical accuracy of the total energy with order $n=3$ and $n_\G = 10$. The error of approach (iii) exceeds 0.1 kcal/mol, with $s_{\min}$ four orders of magnitude larger. This may suggest a computational advantage of less linear dependency. With 0.98 overlap selection threshold, the approach (iii) reaches the subchemical accuracy with a larger $\Delta E$ and smaller $s_{\min}$ than approaches (i) and (ii). This suggests it is better to include the 2s$\rightarrow$1s channel.

The conceptual advantage between approaches (i) and (ii) is less obvious.

The approach (ii) have more similar accuracies to the approach (i), but with fewer dimensions in the same $n$ and $n_\G$. Nevertheless, as discussed in the previous work \cite{wang2026free}, to utilize the computational efficiency of hierarchical decontractions, further simplification of integral evaluations would require the L\"{o}wdin rules \cite{lowdin1955quantum_a,lowdin1955quantum_b}  and its extension \cite{nakashima2020free}, at least factorize the many-electron integrals as finite sum of products of fewer electron integrals. We have not implemented these methods yet.   

Furthermore, at order $n=2$, $\Delta E$ from approach (i) is at the fringe below the subchemical accuracy, whereas approach (ii) is above the subchemical accuracy.  In the next Subsubsection, the approach (i) will be used. 

\subsubsection{Geminal induced expansion length}

In Table \ref{tab4}, we present the results of lithium ground state. The subchemical accuracy can be achieved at $n=2, n_\G= n_\G^{\orb} = n_\G^{\gem} = 10$. In this case, $\Delta E$ is similar to the helium ground state calculation in Table \ref{tab1}. This aligns with the previous statements about unnecessary to increase the order $n$ \cite{nakatsuji2007solving,nakatsuji2018solving}.  Since this $\Delta E$ is also at the fringe below the subchemical accuracy, small deviation caused by varying $n_\G^{\orb}$ and $n_\G^{\gem}$ may exceed the subchemical accuracy. For this target accuracy and in this case, the geminal induced expansion length may become less effective.

\begin{table}[H]
\centering
\setlength{\tabcolsep}{3pt}
\caption{Results of the variational calculations via the Gaussian expanded complement functions, hierarchical decontraction, and geminal induced expansion length for the ground state of the lithium atom. 0.95 and 0.98 are the thresholds for the overlap selections in Algorithm \ref{alg_sel}. Orderings $\prec_g = \prec_V = \prec_{\mathrm{Slater}} = \prec_{\mathrm{STO-}n\G} = \uparrow$ are adopted. $n_{\G}$, $n_{\G}^{\orb}$, and $n_\G^{\gem}$ stand for the expansion length without the geminal function, the expansion length for orbital Slater function with the presence of geminal function, and the expansion length for the geminal Slater function with presence of geminal function. $M^\before$ and $M^\after$ are denoted to the number of the complement functions before and after the overlap selection, respectively.  $s_{\mathrm{min}}$ stands for the minimum eigenvalue of the overlap matrix after the overlap selection. $\Delta E$ stands for the computed electronic energy \cite[p. 43]{szabo1996modern} subtracts the reference energy. The values of $s_{\mathrm{min}}$  and $\Delta E$ are rounded to even \cite{python_format,python_round,goldberg1991every}. The integer in square bracket is denoted to 10 to the power of the given integer \cite{safronova2011correlation}.   }
\begin{threeparttable}
\begin{tabular}{ccccccccccc}
\hline
\multicolumn{5}{c}{} & \multicolumn{3}{c}{0.95} & \multicolumn{3}{c}{0.98} \\
$n$ & $n_\G$  & $n_\G^{\orb}$  &  $n_\G^{\gem}$   & $M^\before$ & $M^\after$  &  $s_{\mathrm{min}}$ & $\Delta E$ & $M^\after$  & $s_{\mathrm{min}}$  &  $\Delta  E$   \\ \hline
1 & 10 & 10 & 10 & 361 & 160 & $4.3[-8]$ & $3.87[-3]$ & 200 & $4.0[-10]$ & $3.62[-3]$ \\
 &  & 8 & 8 & 285 & 134 & $3.1[-7]$ & $3.98[-3]$ & 169 & $2.7[-8]$ & $3.81[-3]$ \\
 &  & 6 & 6 & 225 & 113 & $4.8[-7]$ & $4.99[-3]$ & 148 & $3.1[-8]$ & $4.71[-3]$ \\
 & 14 & 14 & 14 & 673 & 278 & $1.2[-8]$ & $3.63[-3]$ & 366 & $8.1[-10]$ & $3.49[-3]$ \\
 &  & 12 & 12 & 565 & 250 & $7.4[-8]$ & $3.65[-3]$ & 338 & $1.0[-9]$ & $3.48[-3]$ \\
 &  & 10 & 10 & 473 & 226 & $2.7[-8]$ & $3.72[-3]$ & 299 & $1.5[-9]$ & $3.52[-3]$ \\
2 & 10 & 10 & 10 & 6921 & 1172 & $3.9[-10]$ & $1.58[-4]$ & 1963 & $8.5[-13]$\tnote{a} & $9.20[-5]$ \\
 &  & 8 & 8 & 4248 & 970 & $1.0[-9]$ & $2.37[-4]$ & 1474 & $2.3[-12]$ & $1.16[-4]$ \\
 &  & 6 & 6 & 2543 & 735 & $2.3[-9]$ & $4.22[-4]$ & 1118 & $6.7[-12]$ & $2.42[-4]$ \\
 & 14 & 14 & 14 & 17417 & 2655 & $6.3[-11]$ & $4.00[-5]$ & 4493 & $-1.2[-13]$\tnote{b} & $2.14[-5]$ \\
 &  & 12 & 12 & 12160 & 2315 & $2.2[-9]$ & $7.41[-5]$ & 3854 & $5.2[-14]$\tnote{c} & $2.11[-5]$ \\
 &  & 10 & 10 & 8303 & 1945 & $8.5[-10]$ & $9.36[-5]$ & 3088 & $2.2[-14]$\tnote{d} & $3.08[-5]$ \\
3 & 10 & 10 & 10 & 85221 & 3748 & $7.1[-12]$ & $8.72[-5]$ & 8819 & $-1.2[-12]$\tnote{e} & $2.04[-5]$ \\
 &  & 8 & 8 & 40556 & 2556 & $9.5[-11]$ & $7.20[-5]$ & 5380 & $2.9[-14]$\tnote{f} & $3.08[-5]$ \\
 &  & 6 & 6 & 17155 & 1607 & $3.1[-10]$ & $1.32[-4]$ & 3119 & $5.0[-13]$\tnote{g} & $6.85[-5]$ \\
 & 14 & 14 & 14 & 284929 & 8749 & $5.3[-12]$ & $5.42[-6]$ & 21441 & $-4.4[-12]$\tnote{h} & $1.17[-6]$ \\
 &  & 12 & 12 & 165792 & 6663 & $5.5[-11]$ & $7.44[-6]$ & 15629 & $-2.4[-12]$\tnote{i} & $1.15[-6]$ \\
 &  & 10 & 10 & 89855 & 4855 & $4.7[-11]$ & $1.27[-5]$ & 10781 & $-1.1[-12]$\tnote{j} & $2.22[-6]$ \\
\hline
\end{tabular}
\begin{tablenotes}
\item[a-j] The number of complement functions are throwing away by the canonical orthogonalization are 1, 7, 12, 16, 46, 11, 2, 406, 180, and 101, for the superscripts a, b, c, d, e, f, g, h, i, and j, respectively.
\end{tablenotes}
\end{threeparttable}\label{tab4}
\end{table}

Notice lithium ground state has 
\begin{itemize}
    \item [(i)] distinct shell structures by means of 1s and 2s electrons
    \item [(ii)] more than one pair of electrons leads to higher order within one pair function, e.g., $e^{- 2 \gamma_{12} r_{12}}$ and inter-pair terms  $e^{-  \gamma_{12} r_{12} -  \gamma_{13} r_{13}} $
\end{itemize}
These aspects indicate possibilities of more elaborated treatments of geminal induced expansion length than Table \ref{tab4}. Hence,  we present preliminary results in Table \ref{tab5}.

\begin{table}[H]
\centering
\setlength{\tabcolsep}{3pt}
\caption{Results of the variational calculations via the Gaussian expanded complement functions, hierarchical decontraction, and geminal induced expansion length for the ground state of the lithium atom. Orderings $\prec_g = \prec_V = \prec_{\mathrm{Slater}} = \prec_{\mathrm{STO-}n\G} = \uparrow$ are adopted. $n_{\G}$, $n_{\G}^{\orb}$, and $n_\G^{\gem}$ stand for the expansion length without the geminal function, the expansion length for orbital Slater function with the presence of geminal function, and the expansion length for the geminal Slater function with presence of geminal function. $M^\before$ and $M^\after$ are denoted to the number of the complement functions before and after the overlap selection, respectively.  $s_{\mathrm{min}}$ stands for the minimum eigenvalue of the overlap matrix after the overlap selection. $\Delta E$ stands for the computed electronic energy \cite[p. 43]{szabo1996modern} subtracts the reference energy. The values of $s_{\mathrm{min}}$  and $\Delta E$ are rounded to even \cite{python_format,python_round,goldberg1991every}. The integer in square bracket is denoted to 10 to the power of the given integer \cite{safronova2011correlation}.  A denotes to (10, 10, 8), (8, 8, 6). B denotes to  (14, 14, 12), (12, 12, 10), (10, 10, 8).  Power and pair correspond to (power) and (pair) level indications in Section \ref{result}, respectively. The thresholds of throwing functions in the canonical orthogonalization are $1.0 \times 10^{-12}$ and $1.0 \times 10^{-9}$ for $n=2$ and $n=3$, respectively.   }
\begin{threeparttable}
\begin{tabular}{ccccccccc}
\hline
$n$ & $n_\G$ & overlap\_threshold  & $n_\G^{\orb}$  &  $n_\G^{\gem}$  & $M^\before$ & $M^\after$  &  $s_{\mathrm{min}}$ & $\Delta E$   \\ \hline
2 & 10 & 0.95 & A$_{\power}$ & A$_{\power}$ & 4816 & 1065 & $1.4[-9]$ & $1.75[-4]$ \\
 &  &  & A$_{\pair}$ & A$_{\pair}$ & 4880 & 1078 & $1.3[-9]$ & $1.51[-4]$ \\
3 & 16 & 0.98 & B$_{\power}$ & B$_{\power}$ & 163397 & 19325 & $-2.7[-12]$\tnote{a} & $7.52[-7]$ \\
 &  &  & B$_{\pair}$ & B$_{\pair}$ & 168445 & 19512 & $-2.7[-12]$\tnote{b} & $3.07[-7]$ \\
\hline
\end{tabular}
\begin{tablenotes}
\item[a-b] The number of complement functions are throwing away by the canonical orthogonalization are 2861 and 2862, for the superscripts a and b, respectively.
\end{tablenotes}
\end{threeparttable}\label{tab5}
\end{table}

The expansion schemes A and B in Table \ref{tab5} are based on
\begin{itemize}
    \item [(i)] the 2s Slater orbital in the initial complement function being away from the nucleus, thus carries fewer Gaussian expansion length; 
    \item [(ii)] higher level, power or pair, carries fewer expansion lengths.
\end{itemize}
 
According to the data in Table \ref{tab5}, the expansion levels based on the number of distinct electron pairs exhibit lower energies than based on the powers of the  geminal functions. At $n=2, n_\G = 10$ with 0.95 overlap matrix based selection threshold, the error of the total energy with respect to the reference value \cite{wang2012variational} from scheme A$_{\pair}$ is below the subchemical accuracy.

Notice using $1.0 \times 10^{-12}$ threshold of removal functions in the canonical orthogonalization, the energy at $n=3, n_\G = 16$ with 0.98 overlap matrix based selection threshold, scheme B$_{\pair}$ is about $-470$ a.u. Hence, as presented in Table \ref{tab5}, we use $1.0 \times 10^{-9}$ threshold of removal in the canonical orthogonalization. The errors of energies with schemes by  B$_{\power}$ and  B$_{\pair}$ are below $1.0 \times 10^{-6}$ a.u..

Hence, to ameliorate the linear dependence issue, the importance of studying variants of overlap matrix based selection,  mentioned in Subsection \ref{he}, increases. It may also be worthwhile to consider perturbative based selection \cite{bender1969studies,whitten1969configuration,huron1973iterative}, as a lower-cost and intermediate layer between overlap and energy selections \cite{nakashima2020free,nakatsuji2020solving} of the complement functions.

The effectiveness of using lower expansion length in larger principal quantum number orbital and multiple distinct geminal pairs suggests a strategy to improve the computational efficiency. Namely, the computation within to complement functions with fewer geminal functions are performed by the L\"{o}wdin rules \cite{lowdin1955quantum_a,lowdin1955quantum_b} and the extensions \cite{nakashima2020free}, at least factorize the many-electron integrals with exact finite sum of products few-electron integrals. The calculations involved multiple geminal functions shall use lower expansion lengths. This approach to some extent, share similarities  to  composite-related methods, that higher order methods combined with smaller basis sets \cite{mckee1981study,pople1989gaussian,east1993heat}.

\section{Summary and outlook}  \label{summary_outlook}

The present work reports the progress of the free complement method with Gaussian expanded complements for many-electron systems. The infrastructure advances are the main focus. We have formulated and implemented the framework includes permutation structure of $N$-electron system, at least for atomic states with s-type orbitals. This enables us to test the theoretical approaches for a broader range of systems.

To this end, we indicated that the choice of contraction/decontraction should not include additional permutation with respect to the initial wavefunction. We compared three approaches regarding the 2s$\rightarrow$1s channel introduced by the potential operator. We suggested to include this type of complement functions.

The geminal induced expansion length, suggested in the previous work, has been adopted. This indicates a combined strategy to enhance the computational efficiency. Namely, the lower complement functions with fewer pairs of geminal functions are handled by the L\"owdin rules \cite{lowdin1955quantum_a,lowdin1955quantum_b} and its extensions \cite{nakashima2020free}, at least with  facorizations of many-electron integrals into exact finite sum fewer electron integrals. The higher order complement functions with more pairs of geminal functions carry fewer expansion length. The exponential form of the Hylleraas configuration interaction wavefunction \cite{wang2012analytical,nakashima2020free,sims2021exponentially} may be helpful in this direction.

Nevertheless, the numerical results of helium and lithium ground states at $n=2$ and $n_\G = 10$ without lower Gaussian expansion length with the geminal function already are at the fringe below the subchemical accuracy. This leaves fewer rooms to further optimize these computations. Preliminary attempted was made for lithium with more elaborated expansion lengths.

Occasionally, fewer expansion lengths with the presence of geminal functions leads to lower variational energy. This was found more common in smaller threshold of overlap matrix based selection of complement functions. Since the variational calculation is on top of the overlap selection. This suggests further studies for other variant of overlap selection, such as keeping the overlap selected complement functions at low order, then the complement functions from higher order as the new candidate shall be compared.   

Since we have only considered a few conditions among all possibilities of geminal induced expansion length, further research in the landscape of expansion lengths would be of interest.

We discussed the extended symmetries in the integrals in the computations of the lithium ground state. More detailed treatment is planned to be reported with possible improvements.

It is of interest to extend the present approach into more-electron systems. In the meanwhile, it may be worthwhile to look into the determination of the $\gamma$ parameter in the $1-e^{-\gamma r}$ type $g$ function. We may write schematically as $\left[ 1 + c (1-e^{-\gamma r}) \right] \Phi_0$. There are two parameters $c$ and $\gamma$. The common conditions $\gamma_i = Z_i - \zeta_i$ and $\gamma_{ij}=0.5$ can be obtained by $c=1$. The possibilities of using higher order cusp conditions \cite{rassolov1996behavior,tew2008second,  10.1063/1.4816281,10.1063/1.4879266,KUROKAWA201659} to determine the $\gamma$ parameter and if this could benefit the subsequent Gaussian expanded complement function computations may be worthwhile to pursue.

\section*{Appendix}
\setcounter{table}{0}
\renewcommand{\thetable}{A\arabic{table}}

In Tables \ref{taba1} and \ref{taba2}, we present comparisons between different orders described in Algorithms \ref{alg_gen_slater} and \ref{alg_gen_gauss} of Subsections  \ref{gen_slater_function} and \ref{gen_gauss_function}. The ground states of helium atom and lithium atom are used.

In Table \ref{taba1}, the variations of $\Delta E$  are about $6 \times 10^{-5}$ a.u. This is comparable to the effect of enlarging $n_\G$ to 14 or $n$ to 3 in Table \ref{tab1}.  Similar trend holds for lithium in Table \ref{taba2}.

\begin{table}[htbp]
\centering
\caption{Calculated ground state energies of the helium atom with $n=2$, $n_\G=10$, 0.95 threshold of the selection of overlap matrix in Algorithm \ref{alg_sel} with combinations of ordering $\prec_g$, $\prec_V$, $\prec_{\mathrm{Slater}}$, and $\prec_{\mathrm{STO-}n\G}$.  $\uparrow, \downarrow$, and S correspond to ascending, descending, and \texttt{SymPy} order, respectively.  The values of $s_{\mathrm{min}}$  and $\Delta E$ are rounded to even \cite{python_format,python_round,goldberg1991every}. The integer in square bracket is denoted to 10 to the power of the given integer \cite{safronova2011correlation}.  }
\setlength{\tabcolsep}{4pt}
\begin{tabular}{llllcccc}
\hline
$\prec_g$ & $\prec_V$ & $\prec_{\mathrm{Slater}}$ & $\prec_{\mathrm{STO-}n\G}$ & $M^\before$ & $M^\after$ & $s_{\min}$ & $\Delta E$ \\
\hline
$\uparrow$ & $\uparrow$ & S & $\uparrow$ & 196 & 71 & $1.7[-5]$ & $1.52714[-4]$ \\
$\uparrow$ & $\uparrow$ & S & $\downarrow$ & 196 & 72 & $8.6[-6]$ & $2.11083[-4]$ \\
$\uparrow$ & $\uparrow$ & $\uparrow$ & $\uparrow$ & 196 & 71 & $1.7[-5]$ & $1.52714[-4]$ \\
$\uparrow$ & $\uparrow$ & $\uparrow$ & $\downarrow$ & 196 & 72 & $8.6[-6]$ & $2.11083[-4]$ \\
$\uparrow$ & $\uparrow$ & $\downarrow$ & $\uparrow$ & 196 & 71 & $1.7[-5]$ & $1.52714[-4]$ \\
$\uparrow$ & $\uparrow$ & $\downarrow$ & $\downarrow$ & 196 & 72 & $8.6[-6]$ & $2.11083[-4]$ \\
$\uparrow$ & $\downarrow$ & S & $\uparrow$ & 196 & 71 & $1.7[-5]$ & $1.52714[-4]$ \\
$\uparrow$ & $\downarrow$ & S & $\downarrow$ & 196 & 72 & $8.6[-6]$ & $2.11083[-4]$ \\
$\uparrow$ & $\downarrow$ & $\uparrow$ & $\uparrow$ & 196 & 71 & $1.7[-5]$ & $1.52714[-4]$ \\
$\uparrow$ & $\downarrow$ & $\uparrow$ & $\downarrow$ & 196 & 72 & $8.6[-6]$ & $2.11083[-4]$ \\
$\uparrow$ & $\downarrow$ & $\downarrow$ & $\uparrow$ & 196 & 71 & $1.7[-5]$ & $1.52714[-4]$ \\
$\uparrow$ & $\downarrow$ & $\downarrow$ & $\downarrow$ & 196 & 72 & $8.6[-6]$ & $2.11083[-4]$ \\
$\downarrow$ & $\uparrow$ & S & $\uparrow$ & 196 & 72 & $3.8[-6]$ & $1.95626[-4]$ \\
$\downarrow$ & $\uparrow$ & S & $\downarrow$ & 196 & 72 & $3.4[-6]$ & $2.10996[-4]$ \\
$\downarrow$ & $\uparrow$ & $\uparrow$ & $\uparrow$ & 196 & 72 & $3.8[-6]$ & $1.95626[-4]$ \\
$\downarrow$ & $\uparrow$ & $\uparrow$ & $\downarrow$ & 196 & 72 & $3.4[-6]$ & $2.10996[-4]$ \\
$\downarrow$ & $\uparrow$ & $\downarrow$ & $\uparrow$ & 196 & 72 & $3.8[-6]$ & $1.95626[-4]$ \\
$\downarrow$ & $\uparrow$ & $\downarrow$ & $\downarrow$ & 196 & 72 & $3.4[-6]$ & $2.10996[-4]$ \\
$\downarrow$ & $\downarrow$ & S & $\uparrow$ & 196 & 72 & $3.8[-6]$ & $1.95626[-4]$ \\
$\downarrow$ & $\downarrow$ & S & $\downarrow$ & 196 & 72 & $3.4[-6]$ & $2.10996[-4]$ \\
$\downarrow$ & $\downarrow$ & $\uparrow$ & $\uparrow$ & 196 & 72 & $3.8[-6]$ & $1.95626[-4]$ \\
$\downarrow$ & $\downarrow$ & $\uparrow$ & $\downarrow$ & 196 & 72 & $3.4[-6]$ & $2.10996[-4]$ \\
$\downarrow$ & $\downarrow$ & $\downarrow$ & $\uparrow$ & 196 & 72 & $3.8[-6]$ & $1.95626[-4]$ \\
$\downarrow$ & $\downarrow$ & $\downarrow$ & $\downarrow$ & 196 & 72 & $3.4[-6]$ & $2.10996[-4]$ \\
\hline
\end{tabular} \label{taba1}
\end{table}

\begin{table}[htbp]
\centering
\caption{Calculated ground state energies of the lithium atom with $n=2$, $n_\G=10$, 0.95 threshold of the selection of overlap matrix in Algorithm \ref{alg_sel} with combinations of ordering $\prec_g$, $\prec_V$, $\prec_{\mathrm{Slater}}$, and $\prec_{\mathrm{STO-}n\G}$.  $\uparrow, \downarrow$, and S correspond to ascending, descending, and \texttt{SymPy} order, respectively.  The values of $s_{\mathrm{min}}$  and $\Delta E$ are rounded to even \cite{python_format,python_round,goldberg1991every}. The integer in square bracket is denoted to 10 to the power of the given integer \cite{safronova2011correlation}.  }
\setlength{\tabcolsep}{4pt}
\begin{tabular}{llllcccc}
\hline
$\prec_g$ & $\prec_V$ & $\prec_{\mathrm{Slater}}$ & $\prec_{\mathrm{STO-}n\G}$ & $M^\before$ & $M^\after$ & $s_{\min}$ & $\Delta E$ \\
\hline
$\uparrow$ & $\uparrow$ & S & $\uparrow$ & 6921 & 1172 & $3.9[-10]$ & $1.58529[-4]$ \\
$\uparrow$ & $\uparrow$ & S & $\downarrow$ & 6921 & 1165 & $2.0[-11]$ & $1.91833[-4]$ \\
$\uparrow$ & $\uparrow$ & $\uparrow$ & $\uparrow$ & 6921 & 1172 & $3.9[-10]$ & $1.58529[-4]$ \\
$\uparrow$ & $\uparrow$ & $\uparrow$ & $\downarrow$ & 6921 & 1165 & $2.0[-11]$ & $1.91833[-4]$ \\
$\uparrow$ & $\uparrow$ & $\downarrow$ & $\uparrow$ & 6921 & 1173 & $5.8[-9]$ & $1.58209[-4]$ \\
$\uparrow$ & $\uparrow$ & $\downarrow$ & $\downarrow$ & 6921 & 1163 & $1.1[-8]$ & $1.89433[-4]$ \\
$\uparrow$ & $\downarrow$ & S & $\uparrow$ & 6921 & 1172 & $3.9[-10]$ & $1.58529[-4]$ \\
$\uparrow$ & $\downarrow$ & S & $\downarrow$ & 6921 & 1165 & $2.0[-11]$ & $1.91833[-4]$ \\
$\uparrow$ & $\downarrow$ & $\uparrow$ & $\uparrow$ & 6921 & 1172 & $3.9[-10]$ & $1.58529[-4]$ \\
$\uparrow$ & $\downarrow$ & $\uparrow$ & $\downarrow$ & 6921 & 1165 & $2.0[-11]$ & $1.91833[-4]$ \\
$\uparrow$ & $\downarrow$ & $\downarrow$ & $\uparrow$ & 6921 & 1173 & $5.8[-9]$ & $1.58209[-4]$ \\
$\uparrow$ & $\downarrow$ & $\downarrow$ & $\downarrow$ & 6921 & 1163 & $1.1[-8]$ & $1.89433[-4]$ \\
$\downarrow$ & $\uparrow$ & S & $\uparrow$ & 6921 & 1174 & $3.9[-9]$ & $1.95584[-4]$ \\
$\downarrow$ & $\uparrow$ & S & $\downarrow$ & 6921 & 1177 & $3.1[-9]$ & $2.14104[-4]$ \\
$\downarrow$ & $\uparrow$ & $\uparrow$ & $\uparrow$ & 6921 & 1181 & $3.8[-9]$ & $1.97471[-4]$ \\
$\downarrow$ & $\uparrow$ & $\uparrow$ & $\downarrow$ & 6921 & 1185 & $2.7[-9]$ & $2.14564[-4]$ \\
$\downarrow$ & $\uparrow$ & $\downarrow$ & $\uparrow$ & 6921 & 1174 & $3.9[-9]$ & $1.95584[-4]$ \\
$\downarrow$ & $\uparrow$ & $\downarrow$ & $\downarrow$ & 6921 & 1177 & $3.1[-9]$ & $2.14104[-4]$ \\
$\downarrow$ & $\downarrow$ & S & $\uparrow$ & 6921 & 1174 & $3.9[-9]$ & $1.95584[-4]$ \\
$\downarrow$ & $\downarrow$ & S & $\downarrow$ & 6921 & 1177 & $3.1[-9]$ & $2.14104[-4]$ \\
$\downarrow$ & $\downarrow$ & $\uparrow$ & $\uparrow$ & 6921 & 1181 & $3.8[-9]$ & $1.97471[-4]$ \\
$\downarrow$ & $\downarrow$ & $\uparrow$ & $\downarrow$ & 6921 & 1185 & $2.7[-9]$ & $2.14564[-4]$ \\
$\downarrow$ & $\downarrow$ & $\downarrow$ & $\uparrow$ & 6921 & 1174 & $3.9[-9]$ & $1.95584[-4]$ \\
$\downarrow$ & $\downarrow$ & $\downarrow$ & $\downarrow$ & 6921 & 1177 & $3.1[-9]$ & $2.14104[-4]$ \\
\hline
\end{tabular} \label{taba2}
\end{table}

\bibliographystyle{pccp}
\bibliography{topic}

\end{document}


\maketitle

\section{Energies for the tables in the main text with appendix}
In this Section, we supplement the data of the tables in the main text of the present work.

\begin{table}[H]
\centering
\caption{Results of the electronic energies for Table 1 in of the main text. $E$ stands for the computed electronic energy \cite[p. 43]{szabo1996modern}. The values of $E$ are rounded to even \cite{python_format,python_round,goldberg1991every}.    }

\begin{threeparttable}
\begin{tabular}{cccc c c}
\hline
\multicolumn{4}{c}{} & \multicolumn{1}{c}{0.95} & \multicolumn{1}{c}{0.98} \\
$n$ & $n_\G$  & $n_\G^{\orb}$  &  $n_\G^{\gem}$   & $ E$ &  $ E$   \\ \hline
1 & 10 & 10 & 10 & -2.898457724949239 & -2.899361406087346 \\
 &  & 8 & 8 & -2.898512899770349 & -2.899322276275270 \\
 &  & 6 & 6 & -2.898321593544422 & -2.899122430222714 \\
 & 14 & 14 & 14 & -2.899060994103392 & -2.899407395290304 \\
 &  & 12 & 12 & -2.899073936863174 & -2.899399757715666 \\
 &  & 10 & 10 & -2.898966218703954 & -2.899419351552445 \\
2 & 10 & 10 & 10 & -2.903571662606288 & -2.903623655848215 \\
 &  & 8 & 8 & -2.903554772705384 & -2.903626191408640 \\
 &  & 6 & 6 & -2.903445629083568 & -2.903560781729435 \\
 & 14 & 14 & 14 & -2.903624522804272 & -2.903702203939668 \\
 &  & 12 & 12 & -2.903654044152820 & -2.903701600385340 \\
 &  & 10 & 10 & -2.903599066465866 & -2.903688318980522 \\
3 & 10 & 10 & 10 & -2.903664606463857 & -2.903702327608151 \\
 &  & 8 & 8 & -2.903643110012948 & -2.903709287200877 \\
 &  & 6 & 6 & -2.903659466605412 & -2.903683975874348 \\
 & 14 & 14 & 14 & -2.903705008966767 & -2.903721547305261 \\
 &  & 12 & 12 & -2.903704591604558 & -2.903722665568273 \\
 &  & 10 & 10 & -2.903692428930518 & -2.903721379667987 \\
\hline
\end{tabular}
\label{tabs1}
\end{threeparttable}
\end{table}

\begin{table}[H]
\centering
\caption{Results of the electronic energies for Table 2 in of the main text. $E$ stands for the computed electronic energy \cite[p. 43]{szabo1996modern}. The values of $E$ are rounded to even \cite{python_format,python_round,goldberg1991every}.    }

\begin{threeparttable}
\begin{tabular}{cccc c c}
\hline
\multicolumn{4}{c}{} & \multicolumn{1}{c}{0.99} & \multicolumn{1}{c}{0.995} \\
$n$ & $n_\G$  & $n_\G^{\orb}$  &  $n_\G^{\gem}$   & $ E$ &  $ E$   \\ \hline
1 & 10 & 10 & 10 & -2.899351786288768 & -2.899388314554220 \\
 &  & 8 & 8 & -2.899330306577097 & -2.899357450662580 \\
 &  & 6 & 6 & -2.899126443700222 & -2.899126443700222 \\
 & 14 & 14 & 14 & -2.899442827811808 & -2.899462205220591 \\
 &  & 12 & 12 & -2.899447824205081 & -2.899451984741641 \\
 &  & 10 & 10 & -2.899404171077118 & -2.899450515963531 \\
2 & 10 & 10 & 10 & -2.903697132499157 & -2.903699356261013 \\
 &  & 8 & 8 & -2.903691081167680 & -2.903694204735238 \\
 &  & 6 & 6 & -2.903660718842522 & -2.903666495337609 \\
 & 14 & 14 & 14 & -2.903711851167798 & -2.903713535838583 \\
 &  & 12 & 12 & -2.903712241500294 & -2.903712892401075 \\
 &  & 10 & 10 & -2.903711631211592 & -2.903712153142685 \\
3 & 10 & 10 & 10 & -2.903716805197379 & -2.903718016576517 \\
 &  & 8 & 8 & -2.903716379635699 & -2.903716609679463 \\
 &  & 6 & 6 & -2.903707523250945 & -2.903708447495148 \\
 & 14 & 14 & 14 & -2.903723993070152 & -2.903724082861758 \\
 &  & 12 & 12 & -2.903724018630928 & -2.903724050133981 \\
 &  & 10 & 10 & -2.903723775365314 & -2.903723809153795 \\
\hline
\end{tabular}
\label{tabs2}
\end{threeparttable}
\end{table}

\begin{table}[H]
\centering
\caption{
Results of the electronic energies for Table 3 in of the main text. $E$ stands for the computed electronic energy \cite[p. 43]{szabo1996modern}. The values of $E$ are rounded to even \cite{python_format,python_round,goldberg1991every}.  
}
\begin{threeparttable}
\begin{tabular}{c c c c}
\hline
\multicolumn{1}{c}{} & \multicolumn{1}{c}{2s $\rightarrow$1s    } & \multicolumn{1}{c}{ 2s $\rightarrow$1s }  & \multicolumn{1}{c}{ 2s}  \\
\multicolumn{1}{c}{} & \multicolumn{1}{c}{ $\nu, \zeta$  } & \multicolumn{1}{c}{  $\zeta$ }  & \multicolumn{1}{c}{ }  \\
$n$ &  $  E$   & $  E$    &      $ E$  \\ \hline
0 & -7.417793833163408 & -7.417793833163408 & -7.417793833163408 \\
1 & -7.474186639122373 & -7.473646610189427 & -7.473311928975534 \\
2 & -7.477902071905781 & -7.477771570137207 & -7.477367053674150 \\
3 & -7.477973102957232 & -7.477975305216043 & -7.477780635560700 \\
 &  &  & -7.477924777303127\tnote{a} \\
\hline
\end{tabular} \label{tabs3}
\begin{tablenotes}
\item[a] 0.98 threshold of overlap selection
\end{tablenotes}
\end{threeparttable}
\end{table}

\begin{table}[H]
\centering
\caption{Results of the electronic energies for Table 4 in of the main text. $E$ stands for the computed electronic energy \cite[p. 43]{szabo1996modern}. The values of $E$ are rounded to even \cite{python_format,python_round,goldberg1991every}.    }

\begin{threeparttable}
\begin{tabular}{cccc c c}
\hline
\multicolumn{4}{c}{} & \multicolumn{1}{c}{0.95} & \multicolumn{1}{c}{0.98} \\
$n$ & $n_\G$  & $n_\G^{\orb}$  &  $n_\G^{\gem}$   & $ E$ &  $ E$   \\ \hline
1 & 10 & 10 & 10 & -7.474186639122373 & -7.474442811723264 \\
 &  & 8 & 8 & -7.474076607253414 & -7.474250482429055 \\
 &  & 6 & 6 & -7.473068728688641 & -7.473348828750808 \\
 & 14 & 14 & 14 & -7.474431827165025 & -7.474567818766253 \\
 &  & 12 & 12 & -7.474408959649541 & -7.474581295484211 \\
 &  & 10 & 10 & -7.474339322228801 & -7.474541774486924 \\
2 & 10 & 10 & 10 & -7.477902071905781 & -7.477968344343745 \\
 &  & 8 & 8 & -7.477823211540186 & -7.477944334996827 \\
 &  & 6 & 6 & -7.477638357884382 & -7.477818500723236 \\
 & 14 & 14 & 14 & -7.478020300779834 & -7.478038914574475 \\
 &  & 12 & 12 & -7.477986250972759 & -7.478039193105172 \\
 &  & 10 & 10 & -7.477966752026629 & -7.478029566578880 \\
3 & 10 & 10 & 10 & -7.477973102957232 & -7.478039973289177 \\
 &  & 8 & 8 & -7.477988343125418 & -7.478029534148643 \\
 &  & 6 & 6 & -7.477928672831132 & -7.477991820264842 \\
 & 14 & 14 & 14 & -7.478054905018594 & -7.478059151767293 \\
 &  & 12 & 12 & -7.478052884306636 & -7.478059175670778 \\
 &  & 10 & 10 & -7.478047651734233 & -7.478058099842321 \\
\hline
\end{tabular}
\label{tabs4}
\end{threeparttable}
\end{table}

\begin{table}[H]
\centering
\caption{Results of the electronic energies for Table 5 in of the main text. $E$ stands for the computed electronic energy \cite[p. 43]{szabo1996modern}. The values of $E$ are rounded to even \cite{python_format,python_round,goldberg1991every}.    }
\begin{threeparttable}
\begin{tabular}{cccccc}
\hline
$n$ & $n_\G$ & overlap\_threshold  & $n_\G^{\orb}$  &  $n_\G^{\gem}$   & $ E $   \\ \hline
2 & 10 & 0.95 & A$_{\power}$ & A$_{\power}$ & -7.477884876105523 \\
 &  &  & A$_{\pair}$ & A$_{\pair}$ & -7.477909603727512 \\
3 & 16 & 0.98 & B$_{\power}$ & B$_{\power}$ & -7.478059571776074 \\
 &  &  & B$_{\pair}$ & B$_{\pair}$ & -7.478060016747318 \\
\hline
\end{tabular}
\end{threeparttable}\label{tabs5}
\end{table}

\begin{table}[htbp]
\centering
\caption{ Results of the electronic energies for Table A1 in of the main text. $E$ stands for the computed electronic energy \cite[p. 43]{szabo1996modern}. The values of $E$ are rounded to even \cite{python_format,python_round,goldberg1991every}.    }
\begin{tabular}{llllc}
\hline
$\prec_g$ & $\prec_V$ & $\prec_{\mathrm{Slater}}$ & $\prec_{\mathrm{STO-}n\G}$ & $ E$ \\
\hline
$\uparrow$ & $\uparrow$ & S & $\uparrow$ & -2.903571662606280 \\
$\uparrow$ & $\uparrow$ & S & $\downarrow$ & -2.903513294116280 \\
$\uparrow$ & $\uparrow$ & $\uparrow$ & $\uparrow$ & -2.903571662606280 \\
$\uparrow$ & $\uparrow$ & $\uparrow$ & $\downarrow$ & -2.903513294116280 \\
$\uparrow$ & $\uparrow$ & $\downarrow$ & $\uparrow$ & -2.903571662606280 \\
$\uparrow$ & $\uparrow$ & $\downarrow$ & $\downarrow$ & -2.903513294116280 \\
$\uparrow$ & $\downarrow$ & S & $\uparrow$ & -2.903571662606280 \\
$\uparrow$ & $\downarrow$ & S & $\downarrow$ & -2.903513294116280 \\
$\uparrow$ & $\downarrow$ & $\uparrow$ & $\uparrow$ & -2.903571662606280 \\
$\uparrow$ & $\downarrow$ & $\uparrow$ & $\downarrow$ & -2.903513294116280 \\
$\uparrow$ & $\downarrow$ & $\downarrow$ & $\uparrow$ & -2.903571662606280 \\
$\uparrow$ & $\downarrow$ & $\downarrow$ & $\downarrow$ & -2.903513294116280 \\
$\downarrow$ & $\uparrow$ & S & $\uparrow$ & -2.903528750728439 \\
$\downarrow$ & $\uparrow$ & S & $\downarrow$ & -2.903513380623998 \\
$\downarrow$ & $\uparrow$ & $\uparrow$ & $\uparrow$ & -2.903528750728439 \\
$\downarrow$ & $\uparrow$ & $\uparrow$ & $\downarrow$ & -2.903513380623998 \\
$\downarrow$ & $\uparrow$ & $\downarrow$ & $\uparrow$ & -2.903528750728439 \\
$\downarrow$ & $\uparrow$ & $\downarrow$ & $\downarrow$ & -2.903513380623998 \\
$\downarrow$ & $\downarrow$ & S & $\uparrow$ & -2.903528750728439 \\
$\downarrow$ & $\downarrow$ & S & $\downarrow$ & -2.903513380623998 \\
$\downarrow$ & $\downarrow$ & $\uparrow$ & $\uparrow$ & -2.903528750728439 \\
$\downarrow$ & $\downarrow$ & $\uparrow$ & $\downarrow$ & -2.903513380623998 \\
$\downarrow$ & $\downarrow$ & $\downarrow$ & $\uparrow$ & -2.903528750728439 \\
$\downarrow$ & $\downarrow$ & $\downarrow$ & $\downarrow$ & -2.903513380623998 \\
\hline
\end{tabular} \label{taba1}
\end{table}

\begin{table}[htbp]
\centering
\caption{Results of the electronic energies for Table A2 in of the main text. $E$ stands for the computed electronic energy \cite[p. 43]{szabo1996modern}. The values of $E$ are rounded to even \cite{python_format,python_round,goldberg1991every}.    }
\begin{tabular}{llllc}
\hline
$\prec_g$ & $\prec_V$ & $\prec_{\mathrm{Slater}}$ & $\prec_{\mathrm{STO-}n\G}$ & $ E$ \\
\hline
$\uparrow$ & $\uparrow$ & S & $\uparrow$ & -7.477901795141402 \\
$\uparrow$ & $\uparrow$ & S & $\downarrow$ & -7.477868491055026 \\
$\uparrow$ & $\uparrow$ & $\uparrow$ & $\uparrow$ & -7.477901795141402 \\
$\uparrow$ & $\uparrow$ & $\uparrow$ & $\downarrow$ & -7.477868491055026 \\
$\uparrow$ & $\uparrow$ & $\downarrow$ & $\uparrow$ & -7.477902115245417 \\
$\uparrow$ & $\uparrow$ & $\downarrow$ & $\downarrow$ & -7.477870891368807 \\
$\uparrow$ & $\downarrow$ & S & $\uparrow$ & -7.477901795141402 \\
$\uparrow$ & $\downarrow$ & S & $\downarrow$ & -7.477868491055026 \\
$\uparrow$ & $\downarrow$ & $\uparrow$ & $\uparrow$ & -7.477901795141402 \\
$\uparrow$ & $\downarrow$ & $\uparrow$ & $\downarrow$ & -7.477868491055026 \\
$\uparrow$ & $\downarrow$ & $\downarrow$ & $\uparrow$ & -7.477902115245417 \\
$\uparrow$ & $\downarrow$ & $\downarrow$ & $\downarrow$ & -7.477870891368807 \\
$\downarrow$ & $\uparrow$ & S & $\uparrow$ & -7.477864739526184 \\
$\downarrow$ & $\uparrow$ & S & $\downarrow$ & -7.477846220011933 \\
$\downarrow$ & $\uparrow$ & $\uparrow$ & $\uparrow$ & -7.477862852618039 \\
$\downarrow$ & $\uparrow$ & $\uparrow$ & $\downarrow$ & -7.477845760176827 \\
$\downarrow$ & $\uparrow$ & $\downarrow$ & $\uparrow$ & -7.477864739526184 \\
$\downarrow$ & $\uparrow$ & $\downarrow$ & $\downarrow$ & -7.477846220011933 \\
$\downarrow$ & $\downarrow$ & S & $\uparrow$ & -7.477864739526184 \\
$\downarrow$ & $\downarrow$ & S & $\downarrow$ & -7.477846220011933 \\
$\downarrow$ & $\downarrow$ & $\uparrow$ & $\uparrow$ & -7.477862852618039 \\
$\downarrow$ & $\downarrow$ & $\uparrow$ & $\downarrow$ & -7.477845760176827 \\
$\downarrow$ & $\downarrow$ & $\downarrow$ & $\uparrow$ & -7.477864739526184 \\
$\downarrow$ & $\downarrow$ & $\downarrow$ & $\downarrow$ & -7.477846220011933 \\
\hline
\end{tabular} \label{taba2}
\end{table}

\section{Lists of complement functions and data in an example computation}
In this Section, we provide lists of complement functions for the lithium ground state calculation.  They are all generated from order, $n=2$, and expansion length, $n_\G=3$  \cite{stewart1970small}, with orderings  $\prec_g = \prec_V = \prec_{\mathrm{Slater}} =\prec_{\mathrm{STO-}n{\G}}$=$\uparrow$ (ascending).

In Table \ref{tab_li_complement-specs}, the list of complement functions in the form of powers of the $g$ function is provided.

\begin{table}[htbp]
  \centering
  \caption{Li complement functions with $g$ functions, $\phi_I = \mathcal A(\Phi_I \Theta), \Phi_0=r_3e^{-\zeta_1 r_1-\zeta_2 r_2-\zeta_3 r_3}$, with $\zeta_1 = \zeta_2 = 2.68$, $\zeta_3 = 0.635$ \cite{PhysRevA.92.012512}.  $g_q = 1-e^{-\gamma_q r_q}$, where $q=1,2,3,12,13,23$. Here $\gamma_q = Z - \zeta_q, q=1,2,3; \gamma_q = 0.5, q=12,13,23 $ at order $n=2$, $n_\G=3$, $\prec_g = \prec_V = \prec_{\mathrm{Slater}} =\prec_{\mathrm{STO-}n{\G}}$=$\uparrow$.  }
  \label{tab_li_complement-specs}
  \small  \begin{tabular}{clcl}
    \hline
    $I$ & $\Phi_I$ & $I$ & $\Phi_I$ \\
    \hline
    1  & $\Phi_0$ & 19 & $g_{3}\,g_{23}\,\Phi_0/r_3$ \\
    2  & $g_{23}\,\Phi_0$ & 20 & $g_{3}\,g_{12}\,\Phi_0$ \\
    3  & $g_{23}\,\Phi_0/r_3$ & 21 & $g_{3}\,g_{12}\,\Phi_0/r_3$ \\
    4  & $g_{12}\,\Phi_0$ & 22 & $g_{3}^{2}\,\Phi_0$ \\
    5  & $g_{12}\,\Phi_0/r_3$ & 23 & $g_{3}^{2}\,\Phi_0/r_3$ \\
    6  & $g_{3}\,\Phi_0$ & 24 & $g_{2}\,g_{23}\,\Phi_0$ \\
    7  & $g_{3}\,\Phi_0/r_3$ & 25 & $g_{2}\,g_{23}\,\Phi_0/r_3$ \\
    8  & $g_{2}\,\Phi_0$ & 26 & $g_{2}\,g_{13}\,\Phi_0$ \\
    9  & $g_{2}\,\Phi_0/r_3$ & 27 & $g_{2}\,g_{13}\,\Phi_0/r_3$ \\
    10  & $g_{23}^{2}\,\Phi_0$ & 28 & $g_{2}\,g_{12}\,\Phi_0$ \\
    11  & $g_{23}^{2}\,\Phi_0/r_3$ & 29 & $g_{2}\,g_{12}\,\Phi_0/r_3$ \\
    12  & $g_{13}\,g_{23}\,\Phi_0$ & 30 & $g_{2}\,g_{3}\,\Phi_0$ \\
    13  & $g_{13}\,g_{23}\,\Phi_0/r_3$ & 31 & $g_{2}\,g_{3}\,\Phi_0/r_3$ \\
    14  & $g_{12}\,g_{23}\,\Phi_0$ & 32 & $g_{2}^{2}\,\Phi_0$ \\
    15  & $g_{12}\,g_{23}\,\Phi_0/r_3$ & 33 & $g_{2}^{2}\,\Phi_0/r_3$ \\
    16  & $g_{12}^{2}\,\Phi_0$ & 34 & $g_{1}\,g_{2}\,\Phi_0$ \\
    17  & $g_{12}^{2}\,\Phi_0/r_3$ & 35 & $g_{1}\,g_{2}\,\Phi_0/r_3$ \\
    18  & $g_{3}\,g_{23}\,\Phi_0$ &    & \\
    \hline
  \end{tabular}
\end{table}

In Table \ref{tab_li_slater}, the list of complement functions after expanding the powers of $1-e^{-\gamma r}$ type $g$ function is provided.

\begin{table}[htbp]
  \centering
  \caption{Li complement functions with expanded Slater functions from $g$ functions,  $\phi_I = \mathcal A(\Phi_I \Theta), \Phi_0=r_3e^{-\zeta_1 r_1-\zeta_2 r_2-\zeta_3 r_3}$, with $\zeta_1 = \zeta_2 = 2.68$, $\zeta_3 = 0.635$ \cite{PhysRevA.92.012512}.  $g_q = 1-e^{-\gamma_q r_q}$, where $q=1,2,3,12,13,23$. Here $\gamma_q = Z - \zeta_q, q=1,2,3; \gamma_q = 0.5, q=12,13,23 $ at order $n=2$, $n_\G=3$, $\prec_g = \prec_V = \prec_{\mathrm{Slater}} =\prec_{\mathrm{STO-}n{\G}}$=$\uparrow$.  }
  \label{tab_li_slater}
  \small
  \begin{tabular}{clcl}
    \hline
    $I$ & expression & $I$ & expression \\
     \hline
     1  & $\Phi_0$ & 19 & $\mathrm{e}^{-\gamma_3 r_3-\gamma_{23} r_{23}}\,\Phi_0$ \\
    2  & $\mathrm{e}^{-\gamma_{23} r_{23}}\,\Phi_0$ & 20 & $\mathrm{e}^{-\gamma_3 r_3-\gamma_{23} r_{23}}\,\Phi_0/r_3$ \\
    3  & $\Phi_0/r_3$ & 21 & $\mathrm{e}^{-\gamma_3 r_3-\gamma_{12} r_{12}}\,\Phi_0$ \\
    4  & $\mathrm{e}^{-\gamma_{23} r_{23}}\,\Phi_0/r_3$ & 22 & $\mathrm{e}^{-\gamma_3 r_3-\gamma_{12} r_{12}}\,\Phi_0/r_3$ \\
    5  & $\mathrm{e}^{-\gamma_{12} r_{12}}\,\Phi_0$ & 23 & $\mathrm{e}^{-2\gamma_3 r_3}\,\Phi_0$ \\
    6  & $\mathrm{e}^{-\gamma_{12} r_{12}}\,\Phi_0/r_3$ & 24 & $\mathrm{e}^{-2\gamma_3 r_3}\,\Phi_0/r_3$ \\
    7  & $\mathrm{e}^{-\gamma_3 r_3}\,\Phi_0$ & 25 & $\mathrm{e}^{-\gamma_2 r_2-\gamma_{23} r_{23}}\,\Phi_0$ \\
    8  & $\mathrm{e}^{-\gamma_3 r_3}\,\Phi_0/r_3$ & 26 & $\mathrm{e}^{-\gamma_2 r_2-\gamma_{23} r_{23}}\,\Phi_0/r_3$ \\
    9  & $\mathrm{e}^{-\gamma_2 r_2}\,\Phi_0$ & 27 & $\mathrm{e}^{-\gamma_2 r_2-\gamma_{13} r_{13}}\,\Phi_0$ \\
    10  & $\mathrm{e}^{-\gamma_2 r_2}\,\Phi_0/r_3$ & 28 & $\mathrm{e}^{-\gamma_2 r_2-\gamma_{13} r_{13}}\,\Phi_0/r_3$ \\
    11  & $\mathrm{e}^{-2\gamma_{23} r_{23}}\,\Phi_0$ & 29 & $\mathrm{e}^{-\gamma_2 r_2-\gamma_{12} r_{12}}\,\Phi_0$ \\
    12  & $\mathrm{e}^{-2\gamma_{23} r_{23}}\,\Phi_0/r_3$ & 30 & $\mathrm{e}^{-\gamma_2 r_2-\gamma_{12} r_{12}}\,\Phi_0/r_3$ \\
    13  & $\mathrm{e}^{-\gamma_{13} r_{13}-\gamma_{23} r_{23}}\,\Phi_0$ & 31 & $\mathrm{e}^{-\gamma_2 r_2-\gamma_3 r_3}\,\Phi_0$ \\
    14  & $\mathrm{e}^{-\gamma_{13} r_{13}-\gamma_{23} r_{23}}\,\Phi_0/r_3$ & 32 & $\mathrm{e}^{-\gamma_2 r_2-\gamma_3 r_3}\,\Phi_0/r_3$ \\
    15  & $\mathrm{e}^{-\gamma_{12} r_{12}-\gamma_{23} r_{23}}\,\Phi_0$ & 33 & $\mathrm{e}^{-2\gamma_2 r_2}\,\Phi_0$ \\
    16  & $\mathrm{e}^{-\gamma_{12} r_{12}-\gamma_{23} r_{23}}\,\Phi_0/r_3$ & 34 & $\mathrm{e}^{-2\gamma_2 r_2}\,\Phi_0/r_3$ \\
    17  & $\mathrm{e}^{-2\gamma_{12} r_{12}}\,\Phi_0$ & 35 & $\mathrm{e}^{-\gamma_1 r_1-\gamma_2 r_2}\,\Phi_0$ \\
    18  & $\mathrm{e}^{-2\gamma_{12} r_{12}}\,\Phi_0/r_3$ & 36 & $\mathrm{e}^{-\gamma_1 r_1-\gamma_2 r_2}\,\Phi_0/r_3$ \\
    \hline
  \end{tabular}
\end{table}

In the file, \texttt{gauss\_cf\_before.txt}, the list of Gaussian expanded complement functions before the overlap matrix based selection in the main text of the present work, are provided. They are converted in double precision and in the form of \texttt{Formatted\_Gauss\_products} in Algorithm 2 of the main text. Two complement functions are separated there by an empty line. 

The first complement function in  \texttt{gauss\_cf\_before.txt} is 

\noindent 0.78875328669696 0.78875328669696 0.024267420303772 0.0 0.0 0.0  0.0013985289171423798
0.78875328669696 0.78875328669696 0.06321044228854 0.0 0.0 0.0  0.003730223700278493
0.78875328669696 0.78875328669696 1.04095694953355 0.0 0.0 0.0  -0.003066864852600875
0.78875328669696 2.91441075229088 0.024267420303772 0.0 0.0 0.0  0.004487407514762546
0.78875328669696 2.91441075229088 0.06321044228854 0.0 0.0 0.0  0.01196902949892377
0.78875328669696 2.91441075229088 1.04095694953355 0.0 0.0 0.0  -0.009840534734485722
0.78875328669696 15.9999493785216 0.024267420303772 0.0 0.0 0.0  0.004639794781969981
0.78875328669696 15.9999493785216 0.06321044228854 0.0 0.0 0.0  0.012375484159095788
0.78875328669696 15.9999493785216 1.04095694953355 0.0 0.0 0.0  -0.010174708127723326
2.91441075229088 0.78875328669696 0.024267420303772 0.0 0.0 0.0  0.004487407514762546
2.91441075229088 0.78875328669696 0.06321044228854 0.0 0.0 0.0  0.01196902949892377
2.91441075229088 0.78875328669696 1.04095694953355 0.0 0.0 0.0  -0.009840534734485722
2.91441075229088 2.91441075229088 0.024267420303772 0.0 0.0 0.0  0.014398576930888944
2.91441075229088 2.91441075229088 0.06321044228854 0.0 0.0 0.0  0.03840457802447932
2.91441075229088 2.91441075229088 1.04095694953355 0.0 0.0 0.0  -0.031574956352738354
2.91441075229088 15.9999493785216 0.024267420303772 0.0 0.0 0.0  0.014887536264971241
2.91441075229088 15.9999493785216 0.06321044228854 0.0 0.0 0.0  0.03970875391538118
2.91441075229088 15.9999493785216 1.04095694953355 0.0 0.0 0.0  -0.032647206041441404
15.9999493785216 0.78875328669696 0.024267420303772 0.0 0.0 0.0  0.004639794781969981
15.9999493785216 0.78875328669696 0.06321044228854 0.0 0.0 0.0  0.012375484159095788
15.9999493785216 0.78875328669696 1.04095694953355 0.0 0.0 0.0  -0.010174708127723326
15.9999493785216 2.91441075229088 0.024267420303772 0.0 0.0 0.0  0.014887536264971241
15.9999493785216 2.91441075229088 0.06321044228854 0.0 0.0 0.0  0.03970875391538118
15.9999493785216 2.91441075229088 1.04095694953355 0.0 0.0 0.0  -0.032647206041441404
15.9999493785216 15.9999493785216 0.024267420303772 0.0 0.0 0.0  0.015393100103202368
15.9999493785216 15.9999493785216 0.06321044228854 0.0 0.0 0.0  0.04105721813965115
15.9999493785216 15.9999493785216 1.04095694953355 0.0 0.0 0.0  -0.033755868112859404

They are related to $\sum_p c^{(p)} e^{- \alpha_1^{(p)} r_1^2 - \alpha_2^{(p)} r_2^2-\alpha_3^{(p)} r_3^2 - \alpha_{12}^{(p)} r_{12}^2 -\alpha_{13}^{(p)} r_{13}^2 - \alpha_{23}^{(p)} r_{23}^2}$. There are 27 lines that correspond to $3^3$ via STO-3G \cite{stewart1970small} on 3 non-zero orbital Slater exponents and 3 zero geminal Slater exponents, in contracted form. In each line, the first to sixth columns correspond to $\alpha_1^{(p)}, \alpha_2^{(p)}, \alpha_3^{(p)}, \alpha_{12}^{(p)}, \alpha_{13}^{(p)}, \alpha_{23}^{(p)}$. The seventh column correspond to $c^{(p)}$.

As another example,

[empty line]

20.048945256 20.048945256 0.8982484389834 0.0 0.0 0.0  1.0

In this example, [empty line] stands for a separating line without content. This term corresponds to totally decontracted
$1.0 \, e^{- 20.048945256 r_1^2 - 20.048945256 r_2^2-0.8982484389834 r_3^2}$.

In the file, \texttt{gauss\_cf\_after.txt}, the list of Gaussian expanded complement functions after the overlap matrix based selection in the main text of the present work, with 0.95 threshold are provided. The format of the data in \texttt{gauss\_cf\_after.txt} is the same as the data in \texttt{gauss\_cf\_before.txt}.

The ground state energy of the lithium atom for this setting is -7.380690097397304 a.u. Digits are rounded to even.

\bibliographystyle{pccp}
\bibliography{topic}